\documentclass[review,3p,sort&compress,number]{elsarticle}

\usepackage{lineno,hyperref,color,soul}
\usepackage[super]{nth}
\usepackage{natbib}
\usepackage{amsmath,amsthm,amsfonts,amssymb}
\usepackage{graphicx}
\usepackage{multirow}
\usepackage{caption}
\usepackage{subcaption}
\usepackage{array}
\usepackage{tabularx}

\usepackage[capitalise]{cleveref}

\modulolinenumbers[5]

\journal{Journal of Computational Physics}

\begin{document}

\begin{frontmatter}

\title{High-Order, Implicit Time Integration of Discrete, Chaotic Dynamical Systems}


\author{Viktoriya Morozova\fnref{MorozovaFootnote}\corref{mycorrespondingauthor}}
\cortext[mycorrespondingauthor]{Corresponding author}
\ead{vzg5138@psu.edu}
\address{Pennsylvania State University, University Park, PA 16802}

\author{James G. Coder\fnref{CoderFootnote}}
\address{Pennsylvania State University, University Park, PA 16802}

\author{Kevin Holst\fnref{HolstFootnote}}
\address{University of Tennessee, Knoxville, TN 37996}

\fntext[MorozovaFootnote]{Graduate Assistant, Department of Aerospace Engineering}
\fntext[CoderFootnote]{Associate Professor, Department of Aerospace Engineering}
\fntext[HolstFootnote]{Research Assistant Professor, Department of Electrical Engineering and Computer Science}

\begin{abstract}
A wide range of implicit time integration methods, including multi-step, implicit
Runge-Kutta, and Galerkin finite-time element schemes, is evaluated in the context
of chaotic dynamical systems. The schemes are applied to solve the Lorenz equations, 
the equation of motion of a Duffing oscillator, and the Kuramoto-Sivashinsky system,
with the goal of finding the most computationally efficient method that 
results in the least expensive model for a chosen level of accuracy. It is found
that the quasi-period of a chaotic system strongly limits the
time-step size that can be used in the simulations, and all schemes fail
once the time-step size reaches a significant fraction of that period. In these conditions, 
the computational cost per time-step becomes one of the most important factors
determining the efficiency of the schemes. The cheaper, second-order schemes
are shown to have an advantage over the higher-order schemes at large time-step sizes, with one
possible exception being the fourth-order continuous Galerkin scheme. The higher-order schemes become more efficient than the lower-order schemes as accuracy requirements tighten. If going
beyond the second-order is necessary for reasons other than computational
efficiency, the fourth-order methods are shown to perform better than the third-order
ones at all time-step sizes.
\end{abstract}


\end{frontmatter}


\section{Introduction}

Chaotic behaviors may be observed in a wide range of physically relevant, non-linear dynamical systems, such as those used to model weather \cite{Lorenz1963}, animal populations \cite{May_1987}, circuits \cite{Matsumoto_1987,Hasler_1987}, and turbulent fluid flows \cite{VastanoMoser_1991}. Chaotic systems are characterized by the presence of one or more positive Lyapunov exponents that cause nearby trajectories in state space to diverge from each other at an exponential rate. Rather than growing without bound, however, the non-linearities of the system allow the trajectories to remain globally bounded and orbit one or more attractors.  Consequently, careful attention must be given to the properties of the time-integration method used in numerical simulations of chaotic systems. Errors in time integration are equivalent to the solution shifting from one trajectory to another between time-steps. As will be reinforced in this work, the stability properties of the scheme alter the discrete eigenvalues of the system, which in turn affects the shape and strength of the attractors. 



The present study is motivated by scale-resolving simulations of fluid-dynamic
turbulence as governed by the Navier-Stokes equations. Such simulations require
specialized, high-fidelity numerical schemes to achieve accurate and stable solutions.
The overall numerical discretization error includes both spatial and temporal
discretizations. Discretization errors are bound by an equation of the form
$\epsilon \leq C(u) h^p$, where $\epsilon$ is the discretization error, $C(u)$ is a
function of the exact solution being discretized, $h$ is the discretization scale, and
$p$ is the order of accuracy of the numerical method used. This error is typically
reduced by reducing the discretization scale or choosing a method with a higher order of
accuracy. Methods with higher orders of accuracy achieve a lower overall error, but
they typically incur a higher computational cost when compared to lower-order methods
at the same discretization scale. The computational cost can often be mitigated by
increasing the discretization scale, while still achieving the desired error levels.

Much attention is usually given to the spatial discretization operator, which should exhibit minimal artificial dissipation; however, less attention is often given to the time operator. Explicit simulations can be readily made higher-order in time (the classic \nth{4}-order Runge-Kutta method and the low-memory, \nth{3}-order strong-stability-preserving Runge-Kutta method of Gottlieb and Shu \cite{GottliebShu_1998} being commonplace), but the maximum time-step size is limited by their stability bounds. For simulations of the Navier-Stokes equations, this limit is tied to the smallest cell dimension. For high-Reynolds number flows typical of aerodynamic applications, cell sizes in the boundary layer coupled with the stability requirements result in a time-step size that is overly restrictive and often impractical to use on real-world problems due to a large separation in relevant scales. For example, simulating turbulent flow over a commercial aircraft wing requires near-wall grid spacings on the order of 5-10 microns, but a characteristic flow length (i.e. the wing chord) greater than 5 meters. In the context of resolving turbulent eddies, the dominant energy-containing length scales of interest in the simulation are commonly orders of magnitude larger than the near-wall spacing. For this reason, simulation of high-Reynolds number turbulent flows favors the use of implicit time marching, allowing the time-step size to be chosen based on resolving flow features of interest. At a minimum, the implicit scheme should be \textit{A}-stable, with \textit{L}-stability preferred due to the presence of large, negative eigenvalues associated with the viscous operator and small near-wall grid spacing. \footnote{\textit{A}-stability means that the solution of a linear ODE is bounded for all eigenvalues in the left-half plane. \textit{L}-stability is a stronger condition requiring that the stability function approaches zero as $\Delta t \rightarrow \infty$ for eigenvalues in the left-half plane. This property makes \textit{L}-stable methods quickly dampen out rapid transients.}



The implicit schemes used in practice are generally those based on low-order backward-difference formulae, such as BDF1, BDF2, and BDF2OPT \cite{Vatsa2010}, and the prevalence of these schemes in the literature is extensive. These schemes are also referred to as linear multi-step methods because the current time-step is computed from a linear combination of one or more previous time-steps and the current time-step. They are known to be robust and to remain stable even in the presence of poor subiterative convergence of the unsteady residual (a common problem in most Navier-Stokes solvers with practical meshes). The downside of these schemes is that no \textit{A}-stable linear multi-step method can have greater than \nth{2}-order accuracy, an axiom known as the second Dahlquist barrier \cite{Dahlquist1963}.\footnote{The BDF2OPT schemes from \citet{Vatsa2010} use additional time-steps to decrease error, but remain \nth{2}-order accurate.} The methods can introduce an excessive amount of numerical dissipation and dispersion, and the lower order of accuracy can drive the time-step size to values near the explicit stability bounds in order to eliminate this error.

Implicit Runge-Kutta scheme are multi-stage, single-step methods, and thus can bypass the second Dahlquist barrier \cite{Dahlquist1963} and offer stability at arbitrarily high orders of accuracy \cite{Bijl2002,Kennedy2016}. They may be constructed to be \textit{A}- and \textit{L}-stable, lending them well to Navier-Stokes simulations \cite{Bijl2002,Kennedy2016}. Within this general class of schemes, the singly diagonally implicit Runge Kutta methods (SDIRK or ESDIRK, depending on whether or not the first stage is explicit) have garnered the most interest from the fluids community because they can be easily implemented in codes that already use multi-step methods [cf. Refs. \cite{PulliamESDIRK2014,Holst2019,Osusky2012}]. However, these methods come with an increased expense per time-step as each implicit stage requires a unique non-linear solve. The benefits of higher-order time marching for vortex-dominated flows have been established in Refs. \cite{PulliamESDIRK2014,Holst2019}, as temporal discretization errors of lower-order methods such as BDF2 manifest similarly to spatial discretization errors \cite{Holst2019,NishikawaESDIRK2018}. Thus, low-order time marching can negate the benefits of a high-order, low-dissipation spatial scheme. Additionally, the results of Pulliam in Ref. \cite{PulliamESDIRK2014} showed a net cost benefit of \nth{3}- and \nth{4}-order ESDIRK schemes over BDF2 when simulating a convecting isentropic vortex. The higher-order schemes are able to take a sufficiently larger time-step than the lower-order scheme for fixed accuracy to more than offset the increased expense per time-step. Fully implicit Runge-Kutta (FIRK) schemes require specialized solvers and entail a significantly greater computational expense than (E)SDIRK schemes; however, they have been garnering some attention in the form of space-time finite-element and spectral-element methods. \cite{JamesonFIRK2017,Diosady2017,Ekici2020}


For turbulent flows, the trade-off between the added expense of higher-order time accuracy and the time-step size necessary to resolve the chaotic dynamics of turbulence is less obvious. In contrast, the benefits of high-order/high-resolution spatial schemes for increasing accuracy at fixed cost or reducing cost at fixed accuracy are well established. This primary objective of this study is to elucidate the costs and benefits of higher-order, implicit time-marching schemes for simulating chaotic systems.
In particular, we adopt an engineering mindset based on current turbulent-flow simulation practices for which time-step size requirements are determined based on low-order schemes (i.e., BDF2) and investigate if more-expensive, higher-order schemes afford sufficiently larger time-step sizes to justify the additional cost.
Various schemes will be considered, including the backward difference formulae, various (E)SDIRK methods, and Galerkin-based finite-element methods (which themselves may be expressed as FIRK methods). The resolution capabilities of each scheme will be quantified using modified frequency analysis and bandwidth efficiency measures based on those proposed by Lele \cite{Lele1992} for spatial discretization. These schemes will then be applied to the Lorenz system \cite{Lorenz1963}, the Duffing oscillator \cite{Duffing1918}, and the Kuramoto-Sivashinsky equation \cite{Kuramoto1976}. For each dynamical system, the convergence of the Lyapunov spectrum 
will be analyzed with respect to the total computational workload.



\section{Background}

\subsection{Overview of Time-Marching Schemes}

The governing equations for turbulent fluid flows may be written as a hyperbolic conservation law,
\begin{equation}
    \frac{\partial q}{\partial t} + \frac{\partial F_i}{\partial x_i} = 0
\end{equation}
defined on a domain of interest. Contemporary simulation practices are rooted in the idea that steady state solutions, if they exist, should be independent of the choice of time-marching scheme \cite{JamesonJST2017}. It is standard practice to employ a method-of-lines strategy where the spatial derivatives are discretized separately from the time derivative. Thus, the governing equation may be regarded as the solution of a semi-discrete ordinary differential equation in time,
\begin{equation}
    \frac{d f}{d t} = \mathcal{R}\left(f,t\right)
\end{equation}
where $\mathcal{R}(f,t)$ is algebraic and generally non-linear.

In simulations where the steady-state solution is the primary objective, liberties may be taken with the time marching. Low-order schemes with local time-step sizes are commonplace and seek to damp out error in the limit as $t \rightarrow \infty$. Turbulence, however, is an inherently unsteady and chaotic phenomenon, requiring accurate temporal resolution of the energy transfer across a potentially wide range of scales. Choosing the optimal time-marching scheme requires balancing the following qualities of the system and the method:
\begin{enumerate}
    \item Accuracy (i.e. the time-step required to accurately resolve the physics)
    \item Stability (i.e. the maximum permissible time-step size)
    \item Cost (i.e. how expensive each time-step is)
\end{enumerate}
The specific trade-offs are largely a function of the spatial resolution requirements for the flow of interest. For example, free turbulent flows permit the spatial distribution of grid points to be more uniform. The desired time-step size for accuracy may then be close to the stability limit of an explicit, higher-order Runge-Kutta method that is computationally inexpensive. Conversely, an implicit method may well be used with little-to-no concern of the stability limits, but the same time-step size would still be required for accuracy. Thus, the net computational cost of an implicit simulation would be much larger than than of an explicit simulation.

Wall-bounded turbulent flows, on the other hand, require grid-point clustering near the wall to adequately resolve the boundary-layer. The first off-wall cells often have extremely high aspect ratios, thereby introducing numerical stiffness that drive explicit stability limits, whereas the regions of interest for accurately resolving turbulent scales are further from the wall and can tolerate much larger time-steps. The disparity between the explicit stability limits imposed by wall grid spacing requirements and accuracy requirements grows rapidly with the Reynolds number of the simulation \cite{ChoiMoin2012}, and the net computational cost typically favors implicit methods.

\subsection{Multi-Step Implicit Methods}
Multi-step implicit schemes are widely used in Navier-Stokes simulations due in large part to their computational simplicity, requiring only a single implicit solve per global time-step, and robustness towards convergence errors. This group of methods includes the well-known backward difference formulae (BDF). The \nth{1}-order BDF1 scheme, also known as the Backward Euler scheme, is given as,
\begin{equation}
    f^{n+1} = f ^{n} + \Delta t\mathcal{R}\left(f^{n+1},t + \Delta t\right)
\end{equation}
where the superscripts $n$ and $n+1$ denote the solution at $t$ and $t + \Delta t$, respectively. The \nth{2}-order BDF2 scheme is,
\begin{equation}
    \label{Eq: BDF2}
    f^{n+1} = \frac{4}{3} f ^{n} - \frac{1}{3} f^{n-1} + \frac{2 \Delta t}{3}\mathcal{R}\left(f^{n+1},t + \Delta t\right)
\end{equation}
Both the BDF1 and BDF2 schemes are $A$-stable and $L$-stable, making them well-suited for stiff dynamical systems and very appealing for Navier-Stokes simulations at high Reynolds numbers. The BDF1 scheme is also algebraically stable. Higher-order BDF schemes exist, but the second Dahlquist barrier precludes them \textit{A}-stable \cite{Dahlquist1963}.
Nyukhtikov et al. \cite{Nyukhtikov2003} showed that the BDF2 method could be combined with higher-order BDF schemes to produce an \textit{A}-stable \nth{2}-order method with a lower error coefficient. These methods, referred to as BDF2OPT, are not as robust as BDF2 and have a larger storage requirement \cite{Vatsa2010}.

Another scheme that falls in the multi-step implicit class is the \nth{2}-order trapezoidal scheme,
\begin{equation}
    f^{n+1} = f ^{n} + \frac{\Delta t}{2}\left[\mathcal{R}\left(f^{n},t\right) + \mathcal{R}\left(f^{n+1},t + \Delta t\right)\right]
\end{equation}
The trapezoidal scheme is well-known as being the basis of the Crank-Nicolson method for solving the heat equation. While it has the lowest error constant and is \textit{A}-stable \cite{Dahlquist1963}, it is neither \textit{L}-stable nor algebraically stable. 
The lack of \textit{L}-stability is particularly problematic on stiff problems, and Hairer and Wanner \cite{SolvingODE1996} demonstrated this characteristic by comparing BDF1 and trapezoidal solutions of the ODE $y^\prime = -2000(y - \text{cos}\ x)$. For the same time-step size, the BDF1 solution rapidly damps out the initial transient without overshooting, while the trapezoidal solution is under-damped and oscillates around the true value. Nevertheless, the trapezoidal method has seen use in Navier-Stokes simulations, notably for integrating the viscous terms in direct numerical simulations of turbulent channel flows \cite{KMM1987}.


A summary listing the order of accuracy and stability characteristics of the multi-step, implicit methods discussed in this section is given in Table \ref{Table: Multi-step Implicit}.

\begin{table}[!ht]
\centering
\caption{Summary of multi-step implicit methods.}
\label{Table: Multi-step Implicit}
\begin{tabular}{ccccc}
    \hline
    Scheme & Order of Accuracy & $A$-stable & $L$-stable & Alg. stable \\
    \hline
    \hline
    BDF1 (Backward Euler) & \nth{1} & Yes & Yes & Yes \\
    BDF2 & \nth{2} & Yes & Yes & No \\
    Trapezoidal & \nth{2} & Yes & No & No \\
    \hline
\end{tabular}
\end{table}

\subsection{Implicit Runge-Kutta Methods}

\subsubsection{SDIRK}

Diagonally implicit Runge-Kutta (DIRK) methods are distinguished by a lower triangular Butcher
tableau. This characteristic means that the nonlinear system can be solved sequentially stage
by stage because the solution at each stage is only reliant on the current and previous stages.
This is computationally advantageous over fully implicit Runge-Kutta (FIRK) methods, because it
reduces the total operations required from $s^3N^3 + s^2N^2$ for FIRK methods to
$sN^3 + sN^2$ for DIRK methods \cite{Chen_2014}, where $s$ is the number of
stages and $N$ is the size of the solution vector. Singly diagonally implicit Runge-Kutta
(SDIRK) impose the additional constraint that the diagonal elements of the Butcher tableau are
all the same value. The constraint allows the computational cost to reduce further to
$N^3 + sN^2$ because the iteration matrix, $I_N - h \gamma J$, is constant across all stages
\cite{Chen_2014}. The general form of the Butcher tableau for SDIRK schemes is
given in Table \ref{SDIRK_general_tableau}.

SDIRK methods have gained traction within finite-element solvers, such as in Refs.
\cite{Newman2015,Holst2019,Williams2019}. They are appealing for Navier-Stokes simulations
in part because they can be constructed with arbitrarily high accuracy while maintaining both
\textit{A}-stability and \textit{L}-stability and because the form of the Jacobian is the same at each stage. Additionally, they maintain compatibility with standard implicit
solver frameworks because each stage can be solved sequentially.



In this study, we consider the \nth{2}-order-accurate two-stage (SDIRK22), \nth{3}-order-accurate three-stage (SDIRK33), and \nth{4}-order-accurate (SDIRK45) schemes, which are all \textit{A}-stable and \textit{L}-stable \cite{Kennedy2016,Newman2015}. Additionally, we also consider \nth{2}-, \nth{3}-, and \nth{4}-order algebraically stable SDIRK schemes, termed SDIRK22Alg, SDIRK33Alg, and SDIRK44Alg, respectively, as documented in Refs. \cite{Alexander1977,Burrage1982,Williams2019}. A summary of the schemes' properties is given in Table \ref{Table: SDIRK}.

\begin{table}[h!]
	\caption{General form of Butcher tableau for SDIRK methods.}
	\label{SDIRK_general_tableau}
	\renewcommand\arraystretch{1.4}
	\centering
	$
	\begin{array}{c|cccc}
	c_1   & \gamma \\
	c_2   & a_{2,1}      & \gamma \\
	\vdots & \vdots    &      & \ddots \\
	c_n   & a_{n,1} & a_{n,2} & \cdots  & \gamma \\ \hline
	   & b_1    & b_2    & \cdots & b_n
	\end{array}
	$
\end{table}

\begin{table}[!h]
\centering
\caption{Summary of SDIRK methods.}
\label{Table: SDIRK}
\begin{tabular}{cccccc}
    \hline
    Scheme & Stages & Order of Accuracy & $A$-stable & $L$-stable & Alg-stable \\
    \hline
    \hline
    SDIRK22 & 2 & \nth{2} & Yes & Yes & No \\
    SDIRK33 & 3 & \nth{3} & Yes & Yes & No \\
    SDIRK45 & 5 & \nth{4} & Yes & Yes & No \\
    SDIRK22Alg & 2 & \nth{2} & Yes & Yes & Yes \\
    SDIRK33Alg & 3 & \nth{3} & Yes & Yes & Yes \\
    SDIRK44Alg & 4 & \nth{4} & Yes & Yes & Yes \\
    \hline
\end{tabular}
\end{table}

\subsubsection{ESDIRK}
One common modification to SDIRK schemes is to make the first stage explicit.
These schemes are termed ESDIRK, and the general form of their Butcher tableaux is given in Table \ref{ESDIRK_general_tableau}. Three schemes of this type considered in this work are the ESDIRK22, ESDIRK33, and ESDIRK45, all defined in Ref. \cite{Kennedy2016} and summarized in Appendix A. For linear systems, they are each equivalent to the SDIRK schemes that have the same number of implicit stages and same order of accuracy. However, the advantage of ESDIRK methods is that the stage order \footnote{Stage order refers to the accuracy of an individual stage rather than the formal order of accuracy for the entire scheme.} is increased beyond that of SDIRK methods, and their tolerance of errors and/or underconvergence of the implicit residual is improved. A summary of the ESDIRK schemes' properties is given in Table \ref{Table: ESDIRK}.

\begin{table}[h!]
	\caption{General form of Butcher tableau for ESDIRK methods.}
	\label{ESDIRK_general_tableau}
	\renewcommand\arraystretch{1.4}
	\centering
	$
	\begin{array}{c|cccc}
	c_1   & 0 \\
	c_2   & \gamma      & \gamma \\
	\vdots & \vdots    &      & \ddots \\
	c_n   & a_{n,1} & a_{n,2} & \cdots  & \gamma \\ \hline
	   & b_1    & b_2    & \cdots & b_n
	\end{array}
	$
\end{table}

\begin{table}[!h]
\centering
\caption{Summary of ESDIRK methods.}
\label{Table: ESDIRK}
\begin{tabular}{cccccc}
    \hline
    Scheme & Stages & Order of Accuracy & \textit{A}-stable & \textit{L}-stable & Alg-stable \\
    \hline
    \hline
    ESDIRK22 & 3 (2 implicit) & \nth{2} & Yes & Yes & No \\
    ESDIRK33 & 4 (3 implicit) & \nth{3} & Yes & Yes & No \\
    ESDIRK45 & 6 (5 implicit) & \nth{4} & Yes & Yes & No \\
    \hline
\end{tabular}
\end{table}

\subsection{Galerkin Finite-Time Elements}

Various families of temporal integration schemes may be derived by assuming a finite-element
representation of the solution between times $t$ and $t + \Delta t$. The values of $f$ are
stored at discrete locations within the element and interpolated using suitable basis functions
associated with each solution point. A set of discrete, algebraic equations may then be obtained
by using the basis functions as test functions in a weak formulation of the differential
equation,
\begin{equation}
    \int_t^{t + \Delta t} \phi \frac{d f}{d t} dt = \int_t^{t + \Delta t} \phi \mathcal{R}\left(f,t\right) dt
\end{equation}
It is generally beneficial to integrate by parts to allow for treatment of discontinuities
between elements,
\begin{equation}\label{Eq: weak ODE}
    \phi_R f^{n+1} - \phi_L f^n - \int_t^{t + \Delta t} f \frac{ d \phi}{d t} = \int_t^{t + \Delta t} \phi \mathcal{R}\left(f,t\right) dt
\end{equation}
The quantities $\phi_L$ and $\phi_R$ are the values of the basis function at the left ($t$) and
right ($t + \Delta t$) sides of the element. For methods that are continuous by construction,
$f^n$ will also be equal to the value of $f$ on the inside of the element at the left interface.
For discontinuous methods, $f^{n+1}$ is both the interior value at the right interface and the
exact Riemann temporal flux across the interface. The integrals in Eq. \ref{Eq: weak ODE} may be evaluated using any suitable quadrature rules. Of particular interest for this study are collocation methods, which eliminates the explicit need for an interpolation rule and permit superconvergence.

The use of the Galerkin test function offers stability by ensuring that errors are orthogonal to the interpolating polynomial; however, this comes at the increased computational cost of all stages being fully coupled. Thus, Galerkin finite time elements may be expressed as fully implicit Runge-Kutta (FIRK) schemes. Dynamical systems with very large dimensionality, such as discrete representations of the Navier-Stokes equations, may incur a prohibitive cost increase; however, novel implementation strategies such as parallel-in-time approximate factorizations \cite{Ramezanian2017} may alleviate these costs and permit a net benefit in terms of wall-time versus accuracy for solutions.

\subsubsection{Continuous Collocation Methods}

The first class of Galerkin schemes we consider are continuous and based on Gauss-Legendre quadrature, which exactly integrates polynomials up to order $2n-1$ where $n$ is the number of quadrature points and does not include endpoints. With two points, this rule can exactly integrate cubic polynomials. While this would seem to suggest \nth{2}-order accuracy under the Galerkin statement, this scheme is found to be exactly equivalent to a standard Gauss-Legendre method and is thus regarded as superconvergent. The associated Butcher tableau for this scheme, hereafter termed 'CG4' as a \nth{4}-order-accurate continuous Galerkin scheme, is shown in Table \ref{legendre_tableau}.

From this tableau, it can be shown that the scheme is \textit{A}-stable but neither \textit{L}-stable nor algebraically
stable. 

\begin{table}[h!]
	\caption{Butcher tableau for \nth{4}-order continuous Galerkin (CG4) method.}
	\label{legendre_tableau}
	\renewcommand\arraystretch{1.4}
	\centering
	$
	\begin{array}{c|cc}
	\frac{3 - \sqrt{3}}{6}   & \frac{1}{4} & \frac{1}{4} - \frac{\sqrt{3}}{6} \\
	\frac{3 + \sqrt{3}}{6} & \frac{1}{4} + \frac{\sqrt{3}}{6} & \frac{1}{4} \\
	\hline
	              & \frac{1}{2}    &  \frac{1}{2} 
	\end{array}
	$
\end{table}



\subsubsection{Discontinuous Collocation Methods}

The second class of Galerkin schemes under consideration are discontinuous and based on Gauss-Lobatto quadrature, which includes the endpoints of the domain and exactly integrates polynomials up to order $2n - 3$. With three quadrature points, fourth-order accuracy may be attained. This scheme is hereafter referred to as 'DG4', and its tableau is provided in Table \ref{gauss-lobatto_tableau}, from which the scheme is found to be \textit{A}-stable, \textit{L}-stable, and algebraically stable. It is important to note that this scheme is equivalent to the \nth{4}-order-accurate Lobatto IIIC method \cite{NorsettWanner1981}; however, it is important to note that Lobatto IIIC methods were derived using variational analysis rather than Galerkin projections. The equivalence of the discontinuous Galerkin schemes and the Lobatto IIIC methods has been observed for other orders of accuracy, but the authors are presently unaware of any formal proofs connecting the two methods for all orders. An \nth{8}-order scheme of the same form was also considered in initial numerical experiments, but its full description is omitted here for the sake of brevity.

\begin{table}[h!]
	\caption{Butcher tableau for \nth{4}-order discontinuous Galerkin (DG4) method.}
	\label{gauss-lobatto_tableau}
	\renewcommand\arraystretch{1.4}
	\centering
	$
	\begin{array}{c|ccc}
	0   & \frac{1}{6} & -\frac{1}{3} & \frac{1}{6} \\
	\frac{1}{2} & \frac{1}{6} & \frac{5}{12} & \frac{1}{6} \\
	1 & \frac{1}{6} & \frac{2}{3} & \frac{1}{6} \\
	\hline
	              & \frac{1}{6}    &  \frac{2}{3} 
	              & \frac{1}{6}
	\end{array}
	$
\end{table}

\begin{table}[!h]
\centering
\caption{Summary of Galerkin-based collocation methods.}
\label{Table: GalerkinMethods}
\begin{tabular}{cccccc}
    \hline
    Scheme & Stages & Order of Accuracy & \textit{A}-stable & \textit{L}-stable & Alg-stable \\
    \hline
    \hline
    CG4 & 2 & \nth{4} & Yes & No & No \\
    DG4 & 3 & \nth{4} & Yes & Yes & Yes \\
    DG8 & 5 & \nth{8} & Yes & Yes & Yes \\
    \hline
\end{tabular}
\end{table}

\section{Frequency Resolution Analysis}

\begin{figure}[h!]
    \centering
    \begin{subfigure}{0.99\textwidth}
        \centering
        \includegraphics[width=\textwidth]{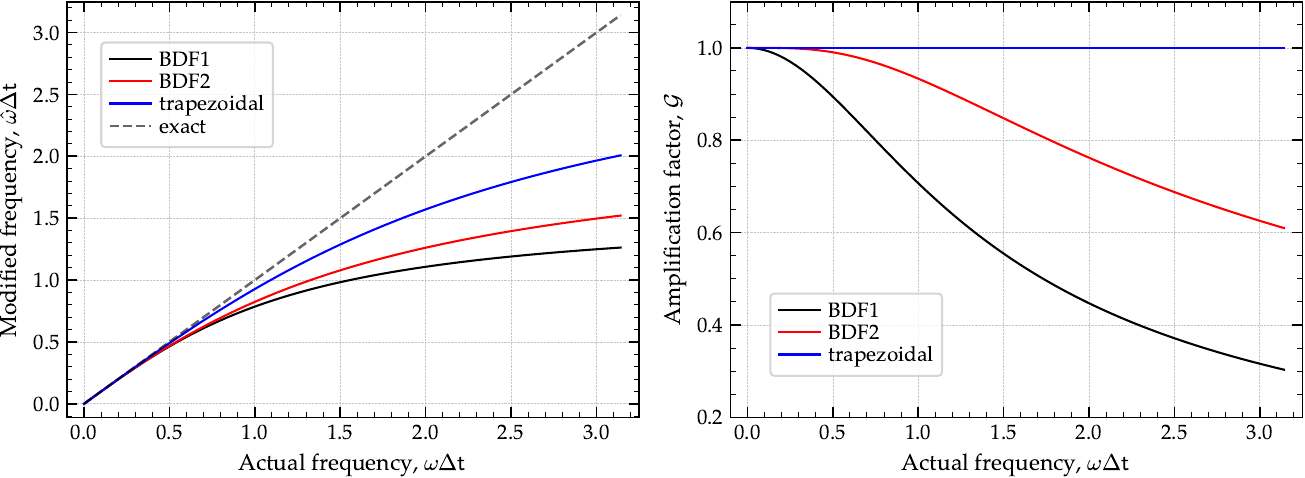}
    
        \label{fig:TKE_production_BDF}
    \end{subfigure}
    

     \caption{Dispersive and dissipative behaviors of various implicit, multi-stage schemes.}
     \label{Fig: BDF errors}
\end{figure}

\begin{figure}[h!]
    \centering
    \begin{subfigure}{0.99\textwidth}
        \centering
        \includegraphics[width=\textwidth]{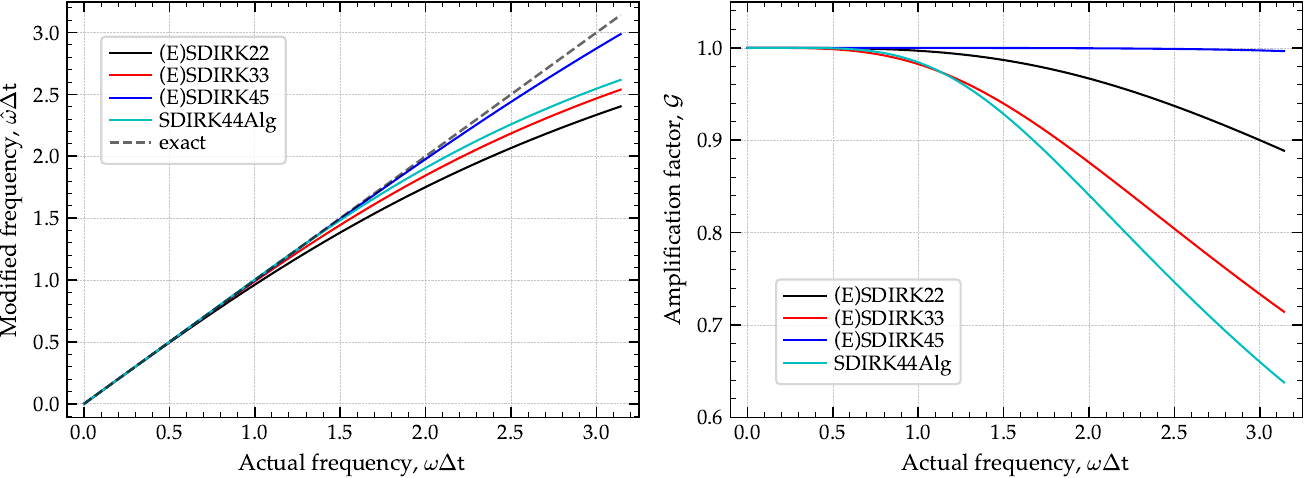}
    
        \label{fig:TKE_production_SDIRK}
    \end{subfigure}
    

     \caption{Dispersive and dissipative behaviors of various diagonally implicit Runge-Kutta schemes.}
     \label{Fig: SDIRK errors}
\end{figure}

\begin{figure}[h!]
    \centering
    \begin{subfigure}{0.99\textwidth}
        \centering
        \includegraphics[width=\textwidth]{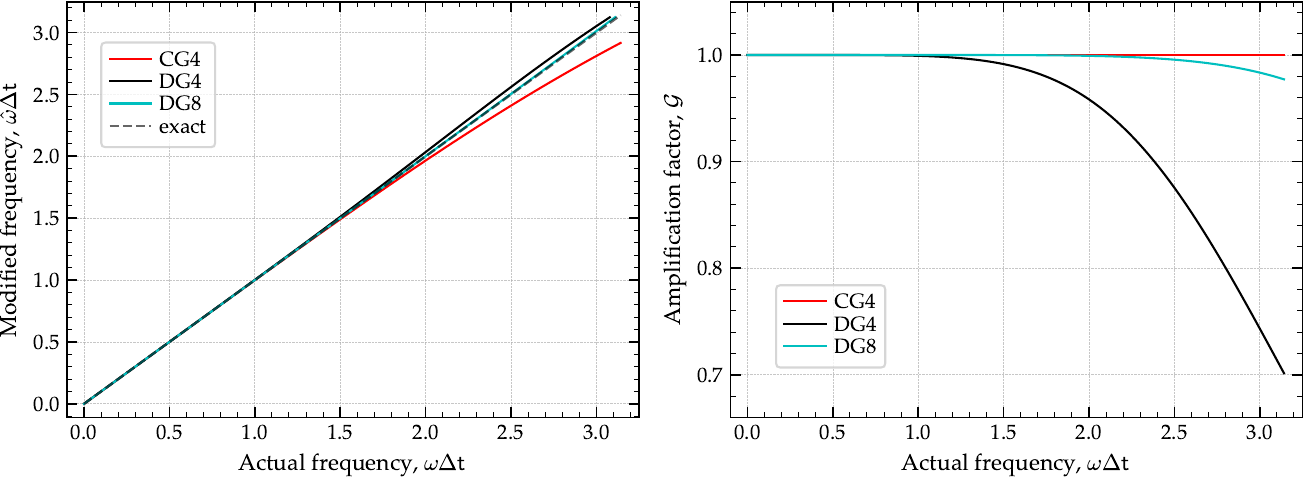}
    
        \label{fig:TKE_production_Galerkin}
    \end{subfigure}
    

     \caption{Dispersive and dissipative behaviors of various Galerkin finite-time-element schemes.}
     \label{Fig: Galerkin errors}
\end{figure}

Analysis of time-marching schemes commonly focuses on their stability characteristics and the leading error term, where even-order schemes are dominated by dispersive error, and odd-order schemes are dominated by diffusive error.
For turbulent flow simulations, it is also important to understand how well these schemes resolve temporal content across the frequency spectrum relative to the Nyquist frequency of the simulation.
To quantify this, we employ a modified frequency analysis that mirrors modified wavenumber analyses used for spatial discretizations \cite{Lele1992}.
Using the linear dynamic system with a purely imaginary eigenvalue given in Eq. \ref{eq:mod_wavenumber_system}, this analysis compares the exact analytic frequency with the discretely-resolved frequency, referred to as the modified frequency. This comparison provides an informative measure of the dispersive and diffusive errors across a range of frequencies.
\begin{equation} \label{eq:mod_wavenumber_system}
    \frac{d z}{d t} = i \omega z
\end{equation}
The dispersive errors may be quantified using the modified frequency, $\tilde{\omega}$, which is defined as
\begin{equation}\label{Eq: Mod. Wavenumber}
    \tilde{\omega} \Delta t = \textnormal{Im} \left[ \ln{ \frac{z \left(t + \Delta t \right)}{z \left( t \right)} }  \right].
\end{equation}
The dissipative errors across the frequency spectrum may be quantified using
\begin{equation}
    \ln{\mathcal{G}} = \textnormal{Re} \left[ \ln{\frac{z \left(t + \Delta t\right)}{z \left(t\right)}} \right],
\end{equation}
where $\mathcal{G}$ represents the amplification factor of the response associated with each frequency $\omega$. Consequently, we may define the ideal scheme as one where $\tilde{\omega} = \omega$ and $\mathcal{G} = 1$ for $0 \leq \omega \Delta t \leq \pi$.
The modified frequency response and amplification ratios for the three groups of schemes (multi-step, (E)SDIRK, and Galerkin) are plotted in Figs. \ref{Fig: BDF errors}, \ref{Fig: SDIRK errors}, and \ref{Fig: Galerkin errors}.

Further following the work by Lele \cite{Lele1992}, we can define a bandwidth efficiency measure as being the percentage of the spectrum that falls within a specified tolerance of relative error,
\begin{equation}\label{Eq: Bandwidth}
    \frac{ \vert \omega - \tilde{\omega} \vert}{\omega} \leq \epsilon_{tol}
\end{equation}
The bandwidth efficiencies for the selected schemes are summarized in Table \ref{Table: Bandwidth} for error tolerances of $10\%$, $1\%$, and $0.1\%$. All (E)SDIRK-type schemes of the same order and same number of stages considered here have identical behaviors for linear problems and are thus combined in the table. 
These results should also be taken in context of the relative workload required for each scheme to advance a single time step. Taking for instance a linear system with an $N \times N$ system matrix, the multi-step schemes all require a single $N \times N$ matrix inversion, the (E)SDIRK schemes require one $N \times N$ inversion per stage, the CG4 scheme requires one $2N \times 2N$ inversion, the DG4 scheme requires one $3N \times 3N$ inversion, and the DG8 scheme requires one $5N \times 5N$ inversion. Thus when normalizing the bandwidth efficiency by the workload, the (E)SDIRK22 method is superior to the BDF2 scheme but not the trapezoidal scheme. For higher acceptable $\epsilon_{tol}$ values, the \nth{2}-order schemes are more efficient than the higher-order schemes, but as the acceptable $\epsilon_{tol}$ decreases, having higher-order accuracy wins out.

\begin{figure}[h!]
  \centering
    \includegraphics[width=0.99\textwidth]{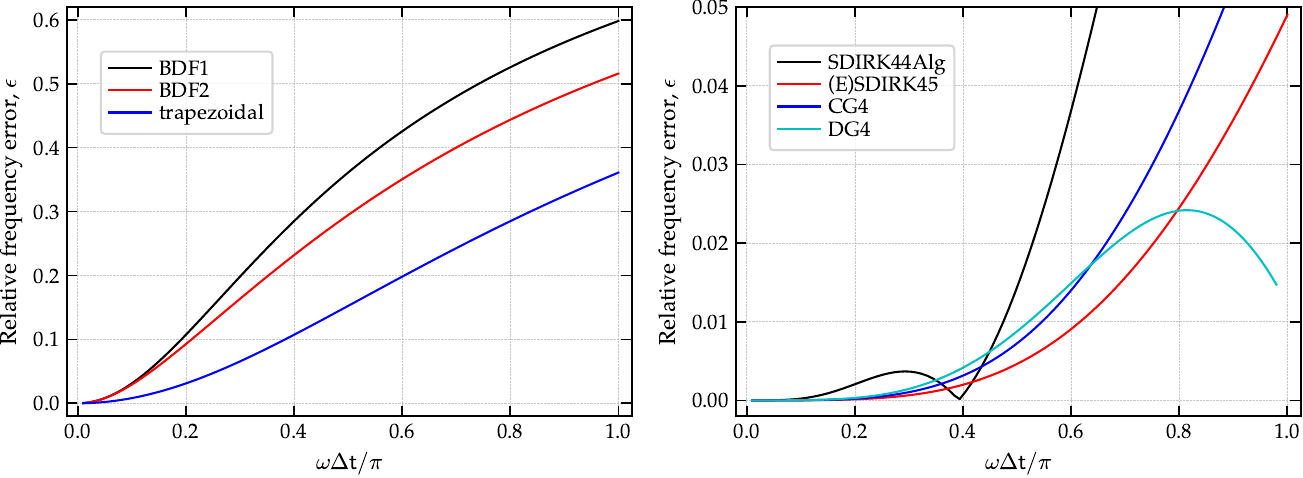}
    \caption{Frequency error for various multi-step (left) and \nth{4}-order multi-stage (right), implicit time-marching schemes. (Note the different scales on the y-axes.)}
    \label{Fig: Freq. Error BDF 4th}
\end{figure}

An interesting, but nevertheless expected result appears for the BDF1 and BDF2 schemes. Despite the higher order of accuracy, BDF2 does not offer significantly greater bandwidth resolution than BDF1, particularly at lower frequencies.
This is depicted in the left panel of Fig. \ref{Fig: Freq. Error BDF 4th}, which compares Eq. \ref{Eq: Bandwidth} for the various multi-step schemes. BDF2 results in significantly less dissipation error than BDF1, as depicted in Fig. \ref{Fig: BDF errors}. The trapezoidal scheme features greater bandwidth resolution than the BDF methods and no dissipation error, as it is a central-difference scheme.


All multi-stage schemes that feature identical numbers of implicit stages and order of accuracy have the same bandwidth efficiencies for all tolerances. This is an expected result as the bandwidth efficiency is a measure of linear performance. The SDIRK44Alg scheme, which has one fewer stage than the other \nth{4}-order (E)SDIRK schemes considered, has remarkably poor performance for lower wavenumbers as suggested by its bandwidth efficiency for $\epsilon_{tol} = 0.001$ being lower than that for the \nth{3}-order algebraically stable scheme. The source of this result is illustrated in the right panel of Fig. \ref{Fig: Freq. Error BDF 4th}, which compares Eq. \ref{Eq: Bandwidth} for the various \nth{4}-order-accurate schemes. The SDIRK44Alg and DG4 schemes show substantially higher dissipation errors than either the CG4 or (E)SDIRK45 schemes. The CG4 scheme is effectively a central difference operator, so it exhibits zero dissipation and sacrifices \textit{L}-stability, while the (E)SDIRK45 scheme has a maximum dissipation error of 0.0035.

\begin{table}[h!]
\centering
\caption{Bandwidth efficiency of various time-marching schemes.}
\label{Table: Bandwidth}
\begin{tabular}{cccc}
    \hline
    Scheme & $\epsilon_{tol}=0.1$ & $\epsilon_{tol}=0.01$ & $\epsilon_{tol}=0.001$  \\
    \hline
    \hline
    BDF1 (Backward Euler) & 0.192 & 0.055 & 0.018 \\
    BDF2 & 0.211 & 0.056 & 0.018 \\
    Trapezoidal & 0.384 & 0.112 & 0.035 \\
    (E)SDIRK22(Alg) & 0.556 & 0.160 & 0.050 \\
    (E)SDIRK33(Alg) & 0.713 & 0.314 & 0.165 \\
    (E)SDIRK45 & 1.00 & 0.617 & 0.336 \\
    SDIRK44Alg & 0.806 & 0.475 & 0.155 \\
    CG4 & 1.00 & 0.547 & 0.298 \\
    DG4 & 1.00 & 0.523 & 0.272 \\
    DG8 & 1.00 & 1.00 & 1.00 \\
    \hline
\end{tabular}
\end{table}

\section{Behaviors for Chaotic Systems}

\subsection{Lorenz System}

The Lorenz system~\cite{Lorenz1963,Sparrow2012}, arguably the most well-known chaotic dynamical system, is a useful test case because of the wealth of studies in the literature into its behavior. This system is described by the set of equations,
\begin{equation}
    \label{Eq: Lorenz}
    \frac{d}{dt} \left[
    \begin{array}{c}
        x \\
        y \\
        z 
    \end{array}
    \right]=
    \left[
    \begin{array}{c}
        \sigma (y - x) \\
        x (\rho - z) - y \\
        xy - \beta z
    \end{array}
    \right]\ .
\end{equation}
The equilibrium points for this system  occur at $\left[x,y,z\right] = \left[ 0,0,0\right]$ and $ \left[ \pm \sqrt{\beta\left(\rho - 1\right)}, \pm \sqrt{\beta\left(\rho - 1\right)}, \rho-1 \right]$, and its Jacobian is expressed as
\begin{equation}
    \label{Eq: Jacobian}
    J =
    \left[
    \begin{array}{ccc}
        -\sigma & \sigma & 0 \\
        \rho - z & -1 & -x \\
        y & x & -\beta
    \end{array}
    \right]\ .
\end{equation}

For this study, we consider the classical parameters of $\sigma = 10$, $\rho = 28$, and $\beta = 8/3$ \cite{Sparrow2012}. The linear stability of the equilibrium points is found by considering the eigenvalues of the system Jacobian. For equilibrium at the origin, there are three real eigenvalues with $\lambda_1 \approx -22.827723$, $\lambda_2 \approx 11.827723$, and $\lambda_3 = -8/3$. Not only does the positive eigenvalue show that this point is linearly unstable, perturbations will grow quite rapidly away from the origin. The other two equilibrium points both possess the set of eigenvalues $\lambda_1 \approx -13.854578$, and $\lambda_{2,3} \approx 0.0939556 \pm 10.194505i$. These points are also linearly unstable since the real part of the complex conjugate pair is positive; however, it is significantly smaller in magnitude than the unstable eigenvalue at the origin.

The linear characteristics of the different schemes provide useful insights into the time-step requirements for simulating the Lorenz system, specifically, the maximum permissible time-step for recovering chaotic dynamics. With a stable, implicit time-marching scheme, the unstable eigenvalue pair $\lambda \approx 0.0939556 \pm 10.194505i$ can be discretely mapped to the left-half plane at moderate time-step sizes, rendering the equilibrium point linearly stable. The maximum $\Delta t$ values for which the equilibrium is linearly unstable is documented in Table \ref{Table: Linear stability}. As expected, the BDF1 scheme shows the most stringent upper bound, requiring a time-step size that is 1-2 orders of magnitude smaller than the other schemes. The BDF2 scheme also shows a stringent bound and is 4 times more restrictive than the \nth{2}-order-accurate SDIRK schemes. The (E)SDIRK45 scheme shows the largest bound; however, the SDIRK44Alg scheme is among the most restrictive despite being the same order of accuracy. It is necessary to note that while ESDIRK45 and SDIRK45 are equivalent for linear problems, SDIRK44Alg is not, due to the different number of stages. Surprisingly, the \nth{3}-order-accurate SDIRK schemes require a time-step size half that of the \nth{2}-order-accurate schemes. The DG4 scheme is more restrictive than \nth{4}-order (E)SDIRK despite the additional complexity associated with a full Butcher tableau. The DG8 scheme is better in this respect; however, it does not show substantial improvement in the stability margin compared to (E)SDIRK45. With the trapezoidal and CG4 schemes, the equilibrium remains unconditionally unstable; however, these schemes lack $L$-stability which can destabilize the overall solution.

\begin{table}[!ht]
\centering
\caption{Maximum permissible step sizes for the Lorenz system based on linear analysis.}
\label{Table: Linear stability}
\begin{tabular}{cc}
    \hline
    Scheme & Maximum $\Delta t$ (Linear Analysis)   \\
    \hline
    \hline
    BDF1 (Backward Euler) & 0.00180794  \\
    BDF2 & 0.03447737  \\
    Trapezoidal & $\infty^{\dagger}$  \\
    (E)SDIRK22 & 0.13735317  \\
    (E)SDIRK33 & 0.07465214  \\
    (E)SDIRK45 & 0.45370034  \\
    SDIRK22Alg & 0.13735317  \\
    SDIRK33Alg & 0.07465214  \\
    SDIRK44Alg & 0.08437928  \\
    CG4 & $\infty^{\dagger}$   \\
    DG4 & 0.16444713   \\
    DG8 & 0.49522675    \\
    \hline 
\end{tabular}\\
\footnotesize{$^{\dagger}$The lack of $L$-stability will cause the stable eigenvalue to become oscillatory and only slightly damped at finite values of $\Delta t$}
\end{table}

To evaluate the non-linear behavior of the time marching schemes, the Lorenz equations were solved numerically using all schemes discussed earlier in this paper except for BDF1, which was deemed incompetitive due to the small maximum permissible time-step size reported in Table~\ref{Table: Linear stability}. Each simulation was run with initial conditions $\left[x,y,z\right] = [1.5, 2.5, 15]$ 
and was simulated to $t = 2 \times 10^4$. The non-linear implicit solutions 
were found using the Newton-Raphson iterative method with a convergence 
criterion based on the $L^2$-norm of the update vector. Unless otherwise 
specified, the convergence tolerance was chosen to be $1 \times 10^{-12}$. 
All cases were run with a constant value of $\Delta t$ (i.e. no time-step adaption) 
to be consistent with the manner in which turbulent fluid flow simulations are typically performed.

\begin{figure}[h!]
    \centering
    \includegraphics[width=0.99\textwidth]{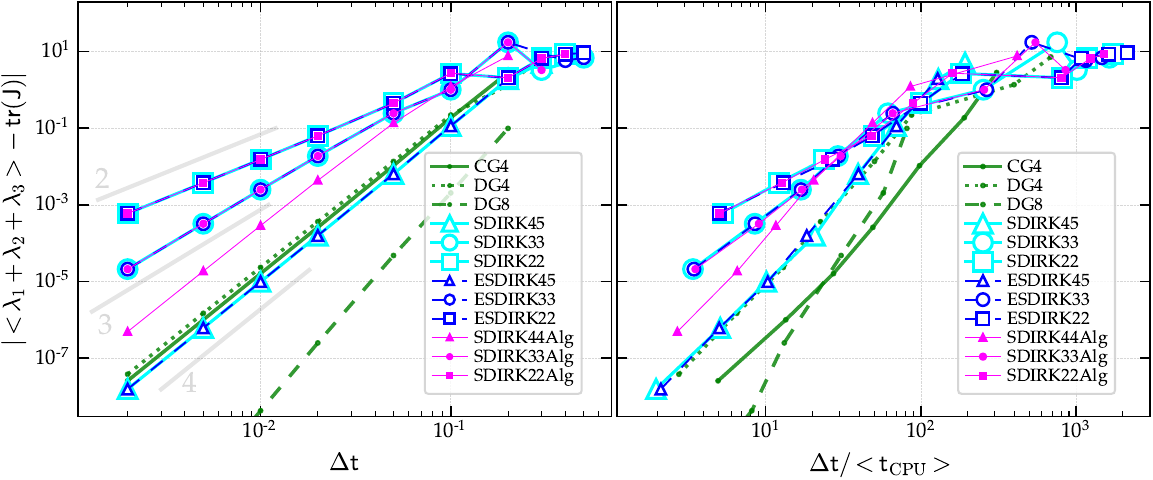}

     \caption{The numerical error as a function of time-step (left panel) and 
inverse workload (right panel) for Lorenz system simulations with different
implicit time integration schemes.}
     \label{Fig: convergence}
\end{figure}

The Lyapunov exponents were computed using a method described in Refs. 
\cite{Benettin1980,Wolf1985}
in which the tangent equation was integrated using the same method as the direct
simulation. A sample Python script showing how the integration was performed 
for the SDIRK45 scheme is provided at \url{https://github.com/vgiryans/sdirk45_lorenz}. The method required a choice of time interval between subsequent
Gram-Schmidt orthonormalizations of the evolved perturbation vectors, and this interval can be larger than the time-step size. 
In this study, we performed
the orthonormalization at every time-step 
and found it sufficient 
to reproduce the exponents previously published in the literature (a discussion of the choice of the orthonormalization interval can be found in \cite{Edson2019}).
However, the same method could not be used for BDF2 scheme, because the
orthonormalization couples different perturbation vectors, and the
vectors orthonormalized at the time-steps $n-1$
and $n$ can no longer be used in the temporal evolution equation (\ref{Eq: BDF2}) to 
find the value of the corresponding perturbation vector at step
$n+1$. Instead, for BDF2 only the leading Lyapunov
exponent was computed, because the orthonormalization of the first perturbation
vector reduces to simply rescaling its magnitude without changing
its direction,
and the rescaling factor for the $(n-1)$-th time-step can be stored in memory and 
used in the $(n+1)$-th time-step.
To allow the initial transients to die off, the Lyapunov exponents were computed
starting from the middle of each simulation. Furthermore, the exponent values
reported here represent their ensemble average value on the interval 
$t = [1.5 \times 10^4, 2 \times 10^4]$, the final quarter of the simulation, which was found to be sufficient to achieve statistical convergence of the exponents.
\footnote{For example, the standard deviations of $\lambda_1$, $\lambda_2$, and 
$\lambda_3$ averaged across
all schemes and time-steps shown further in Fig.~\ref{Fig: convergence} are equal to $6\times 10^{-4}$, $8\times 10^{-4}$,
and $5\times 10^{-3}$, respectively.} 

Verifying numerical convergence of the schemes when working with chaotic
systems is non-trivial, because any small numerical perturbation grows 
into a large difference between the solutions at later times. However, it is
known that the sum of Lyapunov exponents is equal
to the generalized divergence of the flow in phase space \cite{Shin1998}, 
which for the Lorenz system is constant in time and equal to the trace of the
Jacobian, $\mathrm{tr}(J)=-\sigma-\beta-1$. This allows us to use as an error metric the absolute value of the deviation of the averaged sum of 
Lyapunov exponents from the
trace of the Jacobian, $|<\lambda_1+\lambda_2+\lambda_3>-\,\mathrm{tr}(J)|$.

A key component of this study is assessing the relative computational efficiency of the implicit schemes, that is, a measure of the computational effort required
to run a simulation at a specified level of accuracy. The ratio
$\Delta t/<t_{\rm CPU}>$ was used to evaluate the computational efficiency of
these schemes, where $<t_{\rm CPU}>$ is the average CPU time required per
time-step. 
The CPU time per time-step, $t_{\rm CPU}$, is estimated
using the built-in Python timing function, and it includes only the time
needed to update the coordinates of the system, not the time used
to compute the Lyapunov exponents (see the script for the details). The time values
are strictly used for relative comparisons between the schemes, and their absolute
values are not relevant to the discussion. $\Delta t/<t_{\rm CPU}>$ estimated here
reflects the relative amount of numerical operations required by different schemes,
and therefore it can be used for comparison between the schemes.\footnote{An
alternative method for comparing relative workload
could be $N_{\rm sub}/\Delta t$, where
$N_{\rm sub}$ is the number of subiterations used in a scheme, but this metric would
not take into account such factors as size or fullness of the matrix to be inverted,
resulting in a less fair comparison. For example, (E)SDIRK methods have multiple
stages, each of which requires using Newton-Raphson iterative methods, while 
CG4 has only one stage, but the matrix is larger, and the Butcher tableau is full.}
It should also be noted that the implementations of the implicit schemes do not
include any optimizations such as holding the Jacobian constant across stages,
as described in \cite{Chen_2014}, or computing an initial guess to the next stage,
as described in \cite{Kennedy2016}.

Fig.~\ref{Fig: convergence} shows the numerical error as a function of $\Delta t$ (left panel) and
$\Delta t/<t_{\rm CPU}>$ (right panel). 
Time-step sizes of $\Delta t=0.5$, $0.4$, $0.3$, $0.2$, $0.1$, 
$0.05$, $0.02$, $0.01$, $0.005$,
and $0.002$ were tested for all schemes, but some schemes were not
able to reach numerical convergence at the largest time-step sizes, 
and for this reason the corresponding datapoints are absent in the plot.
The left panel of Fig.~\ref{Fig: convergence} shows that for the small time-step sizes all schemes
demonstrate the correct convergence rate, with the highest order schemes
being the most accurate. Different schemes of the same order produce almost
identical errors, apart from SDIRK44Alg which shows poor performance
compared to (E)SDIRK45 and fourth-order Galerkin schemes. As the time-step size
approaches $\sim 0.1$, the convergence rates of the schemes start to deviate from their correct values, and the error becomes large for all schemes.

In the right panel of Fig.~\ref{Fig: convergence}, for a fixed magnitude of the error, the rightmost
scheme is the one that produces the cheapest simulation. 
The plot shows that all second-order schemes 
perform very similarly in terms of the computational efficiency, and the
same is true for all third-order schemes. The fourth-order schemes, on the
other hand, demonstrate a larger spread, with CG4 scheme requiring
less CPU time per time-step than (E)SDIRK45. However, we must mention that the last statement may depend on the hardware and software specifics. The data shown in Fig.~\ref{Fig: convergence}  were generated on a single
CPU core of an HPC cluster, but running the same script on a multicore laptop 
resulted in an opposite picture - the CG4 scheme being slower than 
(E)SDIRK45. We were not able to isolate the reason for poorer performance of CG4 on a laptop, but among the factors that could contribute to this are parallelism of internal Python routines and possibly different algebra packages installed on the two machines. 

The right panel of Fig.~\ref{Fig: convergence} shows that at small time-step sizes, for the values
of error $\lesssim0.1$, the \nth{4}-order schemes are 
computationally more efficient than their lower-order counterparts. 
Therefore, for studies that require reproducing Lyapunov exponents with
 very high accuracy it makes sense to implement the
high-order time marching schemes. At the same time, we
anticipate that there will be many numerical studies of chaotic systems
performed at the largest possible time-step sizes, due to limited
computational resources and/or wall-clock time. For this reason, we are
particularly interested in comparing different time marching algorithms
at large time-step sizes, but Fig.~\ref{Fig: convergence} alone does not let us draw
a clear conclusion about the effectiveness of the schemes in that
regime. On the one hand, the cheapest (rightmost) simulations in the right
panel of Fig.~\ref{Fig: convergence} are represented mostly by the \nth{2}- and \nth{3}-order 
schemes. On the other hand, the
error there becomes so large ($\sim 10$) that it is not clear whether 
the simulations can be trusted at such large time-step sizes.

\begin{figure}[h!]
    \centering
    \includegraphics[width=0.99\textwidth]{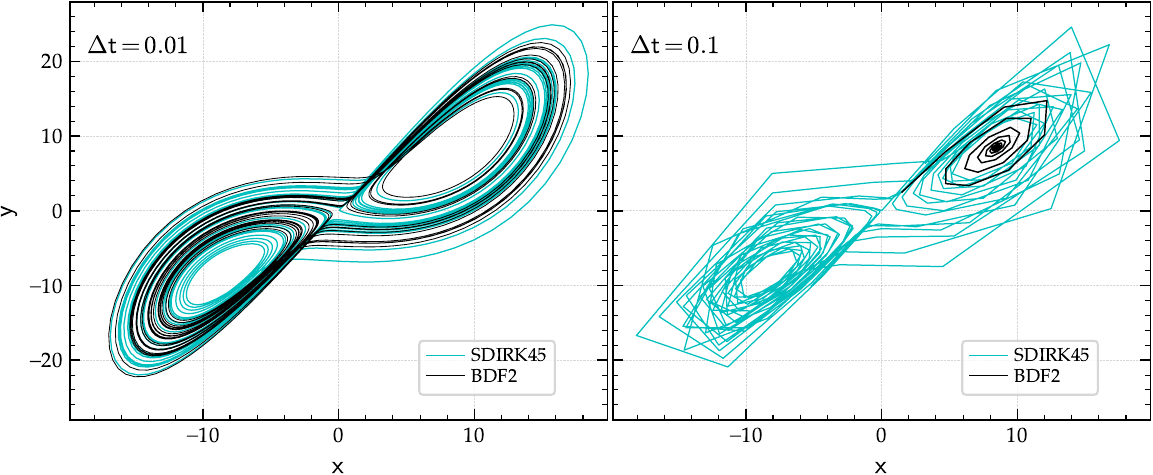}

     \caption{The trajectory of the Lorenz system in $xy$-plane modeled with
SDIRK45 and BDF2 at two different time-step sizes. }
     \label{Fig: trajectories}
\end{figure}

The primary reason for this large numerical error is the fact that
for $\Delta t\gtrsim 0.05$ the time-step size becomes
a noticeable fraction of the period of the system. 
The Lorenz system is chaotic and therefore only
quasi-periodic, but the timescale of its period can be estimated by
looking at the time intervals between the moments when $y$-coordinate 
changes its sign from negative to positive.
These time intervals for the simulation using ESDIRK45 and $\Delta t=0.01$,
range between $0.71$ and $6.4$, with the average of $1.77$. 
This means that at $\Delta t=0.1$ there are on average $\mathcal{O}(10)$ of time-steps per period, which makes it hard to
reproduce the right behavior of the system for any scheme.
To illustrate this point further, the attractor for SDIRK45 and BDF2 for two values of $\Delta t$, $0.01$ and $0.1$, is shown in Fig.~\ref{Fig: trajectories}.
On the left panel ($\Delta t=0.01$), the attractor appears to be represented smoothly
by both schemes, with $\sim100$ points per orbit. 
On the right panel ($\Delta t=0.1$), the time-step size reaches
nearly 10$\%$ of the duration of the orbit, which explains poor performance of all schemes
at large time-step sizes. Nevertheless, SDIRK45 represents the shape of 
the attractor correctly even at $\Delta t=0.1$, while BDF2
quickly converges to the equilibrium point and stays there, in agreement with the
linear analysis of the schemes performed earlier.

\begin{figure}[h]
    \centering
    \includegraphics[width=0.99\textwidth]{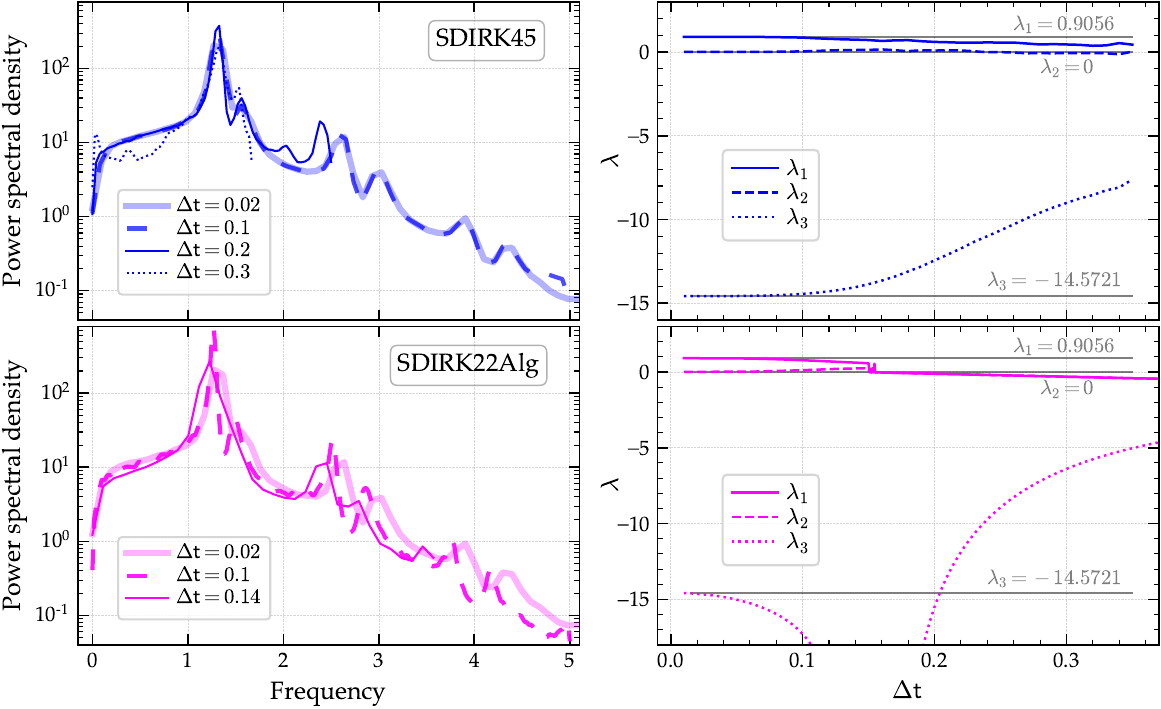}

     \caption{Left panel: Power spectral density of the values of $z$ for the Lorenz system, obtained with SDIRK45 (top) and SDIRK22Alg (bottom) at different values of the time-step size. Right panel: All three Lyapunov exponents obtained with SDIRK45 (top) and SDIRK22Alg (bottom) as a function of $\Delta t$.}
     \label{Fig: spectra Lorenz}
\end{figure}

In addition, while $|<\lambda_1+\lambda_2+\lambda_3>-\,\mathrm{tr}(J)|$ represents a robust measure of the numerical error,
it does not give a full picture of the correctness of individual exponents.
To demonstrate this, Fig.~\ref{Fig: spectra Lorenz} focuses on the performance of SDIRK45 and SDIRK22Alg
at large values of $\Delta t$. The left panel of Fig.~\ref{Fig: spectra Lorenz} shows the power
spectral density of the values of variable $z$ for the different time-step sizes. At $\Delta t=0.02$, the spectrum obtained with either scheme
agrees well with the one published previously in \citet{Farmer1980}. The accuracy
declines with the size of the time-step, but even at the verge of losing the chaotic behavior ($\Delta t = 0.3$ for 
SDIRK45, $\Delta t = 0.14$ for SDIRK22Alg) both schemes are able to robustly reproduce the main peak. The right panel
of Fig.~\ref{Fig: spectra Lorenz} shows all three Lyapunov exponents predicted by these schemes
at large time-step sizes. The values of exponents collected from \citet{Sparrow2012} are shown by
horizontal lines for comparison. The plot shows that at large values of
$\Delta t$ the error in $\lambda_3$ dominates the error in 
$\lambda_1$ by about an order of magnitude, and it is this error
that is mainly responsible for large deviation of $<\lambda_1+\lambda_2+\lambda_3>$ from $\mathrm{tr}(J)$ seen
in Fig.~\ref{Fig: convergence}. Since chaotic properties are determined by the positive Lyapunov exponent, Fig.~\ref{Fig: spectra Lorenz} shows that both SDIRK45 and SDIRK22Alg are better in reproducing the chaotic behavior of the Lorenz system at large time-step sizes than could be inferred from Fig.~\ref{Fig: convergence} alone.

\begin{figure}[h!]
    \centering
    \includegraphics[width=0.99\textwidth]{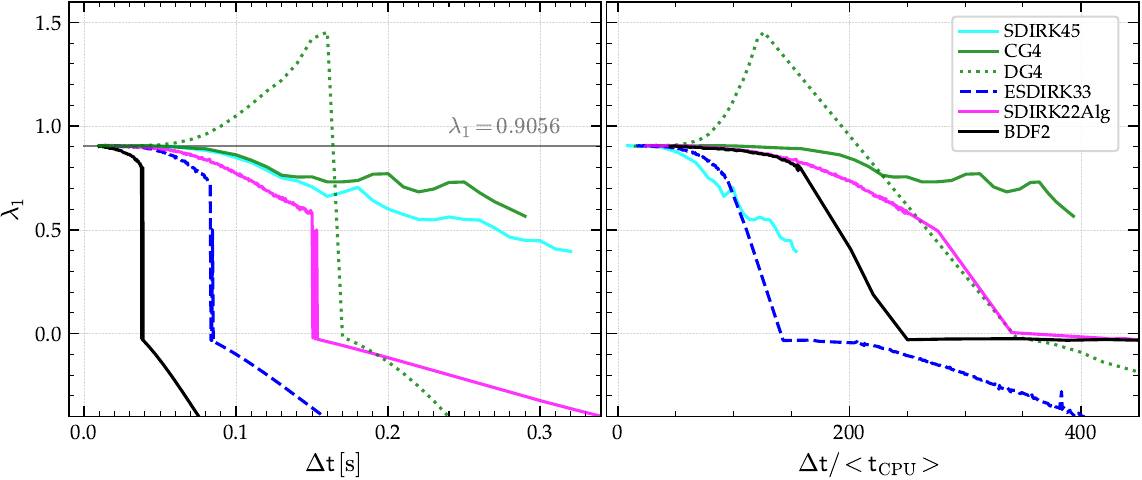}

     \caption{The leading Lyapunov exponent of the Lorenz system as a function
of $\Delta t$ (left panel) and $\Delta t/<t_{\rm CPU}>$ (right panel). }
     \label{Fig: lambdas}
\end{figure}

In Fig.~\ref{Fig: lambdas}, only the leading Lyapunov
exponent, $\lambda_1$, is shown as a function of the time-step
size (left panel) and the inverse workload (right panel). In this figure,
only one SDIRK-type scheme per order is shown, because other schemes of the
same order perform nearly identically (SDIRK44Alg notwithstanding).
The accepted value of $\lambda_1$ \cite{Sparrow2012,Geurts2019} is shown by a
horizontal line in the right panel.
In good agreement with Table~\ref{Table: Linear stability}, ESDIRK33
predicts $\lambda_1$ to become negative for the time-step size $\approx 0.08$,
while SDIRK22Alg fails to predict chaotic behavior of the system at larger time-step size, $\Delta t\approx 0.15$.
This, combined with the smaller number of stages, makes the second-order
scheme approximately two times more efficient at producing chaotic behavior, 
regardless of accuracy, than the third-order scheme, as shown in the right 
panel of Fig.~\ref{Fig: lambdas}.
On the other hand, even though SDIRK45
is able to predict the chaotic behavior of the system up to the largest time-step
size, it is less computationally efficient than the lower-order schemes. CG4 seems to 
outperform all other schemes in terms of efficiency in the
right panel of Fig.~\ref{Fig: lambdas}, but this result is sensitive to the implementation details, and the scheme
lacks \textit{L}-stability that may be needed in some applications
\footnote{Since the implementation of all SDIRK-type schemes in our scripts is very
similar, with differences only in the numbers of stages and the
entries of the Butcher tableaux, the relative performance between the SDIRK
schemes of different order is represented robustly in Figs.~\ref{Fig: convergence} and~\ref{Fig: lambdas}. }. BDF2
starts to fail at the smallest values of the time-step size in the left panel 
of Fig.~\ref{Fig: lambdas}, but due to its low cost per time step it still demonstrates good computational efficiency in the right panel.
DG4 predicts $\lambda_1$ to be positive up to the time-step size of $\approx0.17$,
but the predicted value deviates from the correct value even at the values of
$\Delta t/<t_{\rm CPU}>$ as low as $\sim 100$. 

The results presented in Fig.~\ref{Fig: lambdas} demonstrate that second-order implicit schemes are more efficient at modeling chaotic dynamical systems than higher-order schemes as time-step size and error tolerance increases. When error tolerances are tighter, the use of higher-order schemes results in more efficient simulations. Since using large time-step sizes is one of the main benefits of the time marching schemes, as the error tolerance tightens it may become more efficient to use cheaper explicit schemes. In that case, the ratio between the implicit time-step size required for a chosen accuracy and the maximum possible explicit time-step size must be compared to the ratio between the computational costs of an implicit and explicit time steps. However, from the results presented here it is clear that no implicit scheme can increase the time-step size beyond $\sim 10\%$ of the quasi-period of the system, and using higher-order schemes when running at large time-step sizes would lead to a larger cost of the same quality simulation.

\subsection{Duffing Oscillator}

The Duffing equation, first introduced in \cite{Duffing1918}, describes the 
behavior of a nonlinear driven damped oscillator (for the review of its many
fundamental properties, see \cite{Ueda2004,Guckenheimer1984,Wiggins1988}).
While there are multiple versions of the Duffing equation
that can be found in the literature, here we focus on one of the most thoroughly 
studied version of this equation \cite{Hayashi1953,Ueda1979,Zeni1995}\ ,
\begin{equation}
\label{duf}
\frac{d^2 x}{dt^2} + \delta\frac{dx}{dt} + x^3 = \gamma \cos t\ .
\end{equation}
Here, $x$ is the displacement of the oscillator from its
equilibrium position, $t$ is time, the second term in the left hand side represents
the damping force, while the third term is responsible for the nonlinear behavior.
A term linear in $x$, which was present in the original work by Duffing, is omitted
here, but the studies including that term can be found, among other works, 
in \cite{Holmes1979,Parlitz1986,Brennan2008,Monroe2012}.
The right hand side of Eq.~(\ref{duf}) represents a periodic driving force with the angular
frequency equal to 1, the value commonly used in
the literature.
The parameter space of the Duffing oscillator was studied for a very wide range of the
values of the damping constant, $\delta$, and the amplitude of the driving force,
$\gamma$ \cite{Ueda1980,Bonatto2008,ByattSmith1986,Robinson1989}.
For example, \citet{Bonatto2008} describe self-similar patterns of 
chaotic and periodic regions of the parameter space 
that repeat regularly as the amplitude of the driving force
increases.

\begin{figure}[h!]
    \centering
    \includegraphics[width=0.99\textwidth]{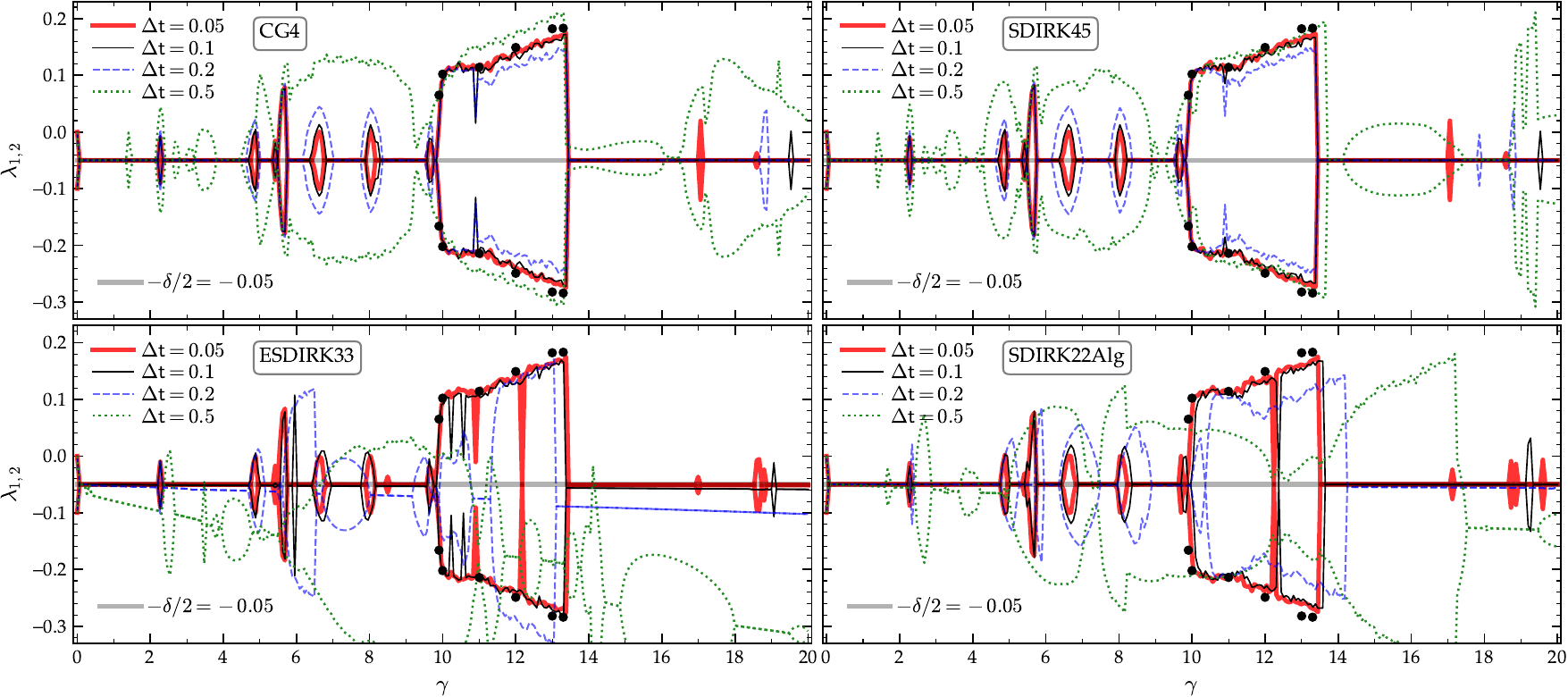}

     \caption{The Lyapunov exponents of the Duffing oscillator for $\delta=0.1$ and
a range of values of the driving force amplitude $\gamma$. }
     \label{Fig: lambdas duffing}
\end{figure}

To reproduce the results obtained in \cite{Ueda1979} and \cite{Zeni1995} on the
dependence of the Lyapunov exponents of the Duffing oscillator on the driving force amplitude, 
we fix the value of $\delta$ to $0.1$ and vary $\gamma$
in the range between $0$ and $20$. Fig.~\ref{Fig: lambdas duffing} shows the Lyapunov exponents
obtained with the different implicit schemes for the time-step sizes 
$\Delta t = 0.05$, $0.1$, $0.2$, and $0.5$. In the previous section it has been shown that
SDIRK-type schemes of the same order perform very similarly,
and in this section we pick only one of those schemes per order,
focusing on SDIRK45, ESDIRK33, and SDIRK22Alg. 
DG4 and DG8 cannot compete with the other schemes
in terms of efficiency because of the large CPU time required per time-step, and for this reason they are not shown here. At the same time, CG4 is used because it has demonstrated some of the best results for the Lorenz
system. BDF2 is not represented in Fig.~\ref{Fig: lambdas duffing}, 
but the simulations of Duffing oscillator performed with BDF2 will be shown in 
the subsequent plots.

The duration of all simulations shown in 
Fig.~\ref{Fig: lambdas duffing} and in the rest of this section is $2\times 10^4$, and only the second half of each
simulation is used to compute the Lyapunov exponents. 
This second half corresponds to over $3000$ cycles of the driving force, which is 
sufficient for the exponents to converge to their final values
(\citet{Zeni1995} demonstrated a good convergence already at $400$ cycles),
while disregarding the first half is more than sufficient to guarantee that the values of the
exponents are not affected by the transients. 
The black markers in Fig.~\ref{Fig: lambdas duffing} indicate the values obtained
in \cite{Ueda1979}, and the simulations presented here agree well with those results at the finest time-step sizes.

Fig.~\ref{Fig: lambdas duffing} shows that increasing the time-step size strongly affects the periodic
regions of the parameter space, where both exponents have to be 
equal to $-0.05$. For example, in the top left panel for CG4, at $\Delta t = 0.5$, 
there are large chaotic regions at the values
of $\gamma$ between  5 and 9, as well as between 16.5 and 20, which should not be
chaotic. At the same time, both \nth{4}-order schemes robustly reproduce
the main chaotic region between the values of $\gamma=10$ and $13.3$ 
for all time-step sizes shown in the figure. The lower order schemes cannot
correctly reproduce that region at the large time-step sizes, with ESDIRK33 being 
less accurate than SDIRK22Alg.

\begin{figure}[h!]
    \centering
\includegraphics[width=0.99\textwidth]{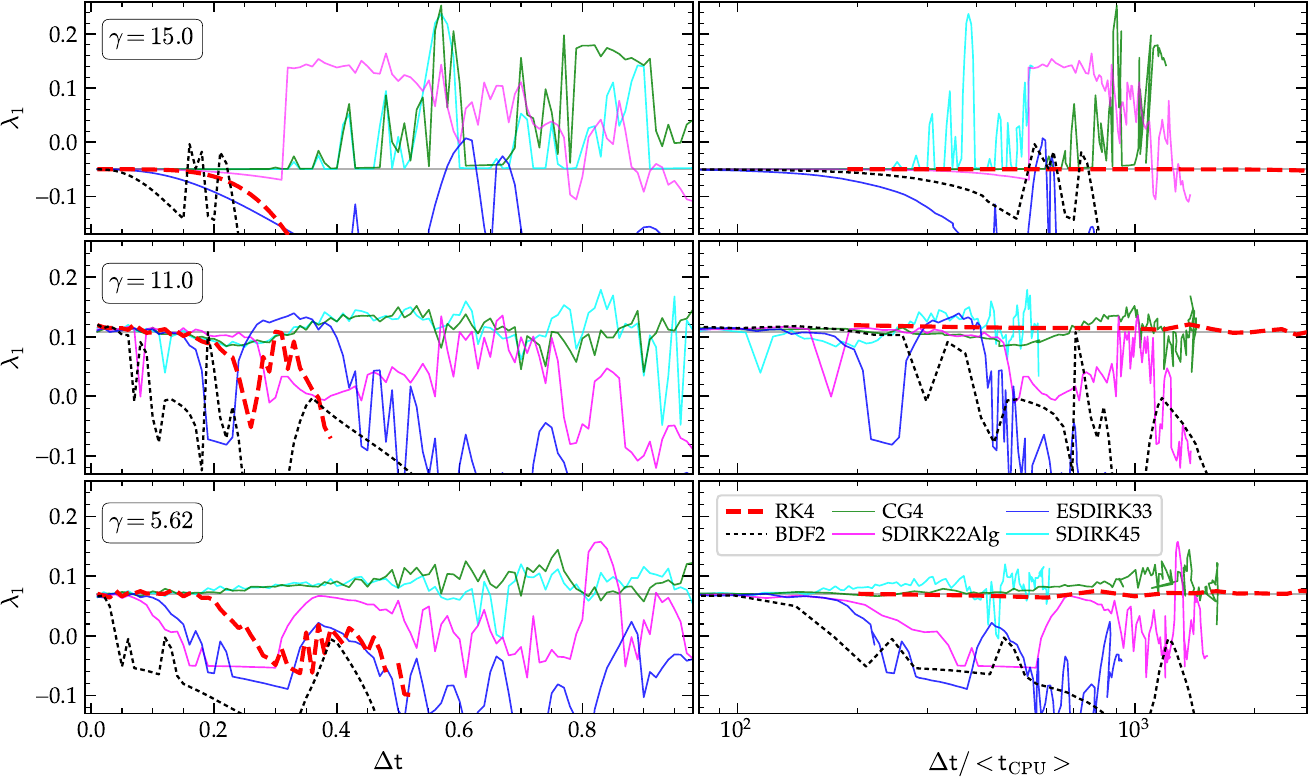}

     \caption{The leading Lyapunov exponent of the Duffing oscillator as a
function of $\Delta t$ (the left column) and $\Delta t/<t_{\rm CPU}>$ (the right
column). }
     \label{Fig: lambdas vs time-step}
\end{figure}

Fig.~\ref{Fig: lambdas vs time-step} further focuses on the dependence of the leading Lyapunov exponent, $\lambda_1$, on the time-step size, looking at the three distinct regions of the parameter space from Fig.~\ref{Fig: lambdas duffing}. The top
row of Fig.~\ref{Fig: lambdas vs time-step} corresponds to $\gamma=15$, where the correct behavior of the system
is periodic. The middle row, $\gamma=11$, represents the main chaotic region
from Fig.~\ref{Fig: lambdas duffing}, while the bottom row, $\gamma=5.62$, represents a small
standalone chaotic region. There is a large number of such regions in Fig.~\ref{Fig: lambdas duffing},
but the one at $\gamma=5.62$ persists at small time-steps, 
and it is also seen in Ref. \cite{Zeni1995}, so it appears to be 
a true region of chaos and not a numerical artifact. The
left column of Fig.~\ref{Fig: lambdas vs time-step} shows $\lambda_1$ as a function of time-step size,
while the right column shows the same values of $\lambda_1$ as a function of
$\Delta t/<t_{\rm CPU}>$, therefore speaking of the computational efficiency
of the schemes. For comparison, we show the results 
obtained with the explicit fourth order Runge-Kutta scheme (RK4).

There is a number of qualitative conclusions that can be drawn from Fig.~\ref{Fig: lambdas vs time-step}.
The left column of Fig.~\ref{Fig: lambdas vs time-step} shows that the two \nth{4}-order
schemes, CG4 and SDIRK45, are much more robust at reproducing
the chaotic behavior of the system (middle and bottom panels) 
than at reproducing its  
periodic behavior (top panel). 
None of the schemes correctly reproduces the periodic
behavior of the system at $\gamma=15$ for the time-step sizes 
above $\approx0.3$ (the top left panel). For
$\gamma=11$ and $5.62$  (the middle left and the bottom left panels), 
both CG4 and SDIRK45 schemes predict the
positive values of $\lambda_1$ up to the time-step sizes of $\Delta t \approx1$,
which is about 4 times larger than the maximum time-step of RK4.
However, even the best implicit schemes cannot achieve a better 
gain in terms of the time-step size, because at this point the time-step 
size reaches a 
significant fraction of the driving force period, $2\pi$. 

\begin{figure}[h!]
    \centering
    \includegraphics[width=0.94\textwidth]{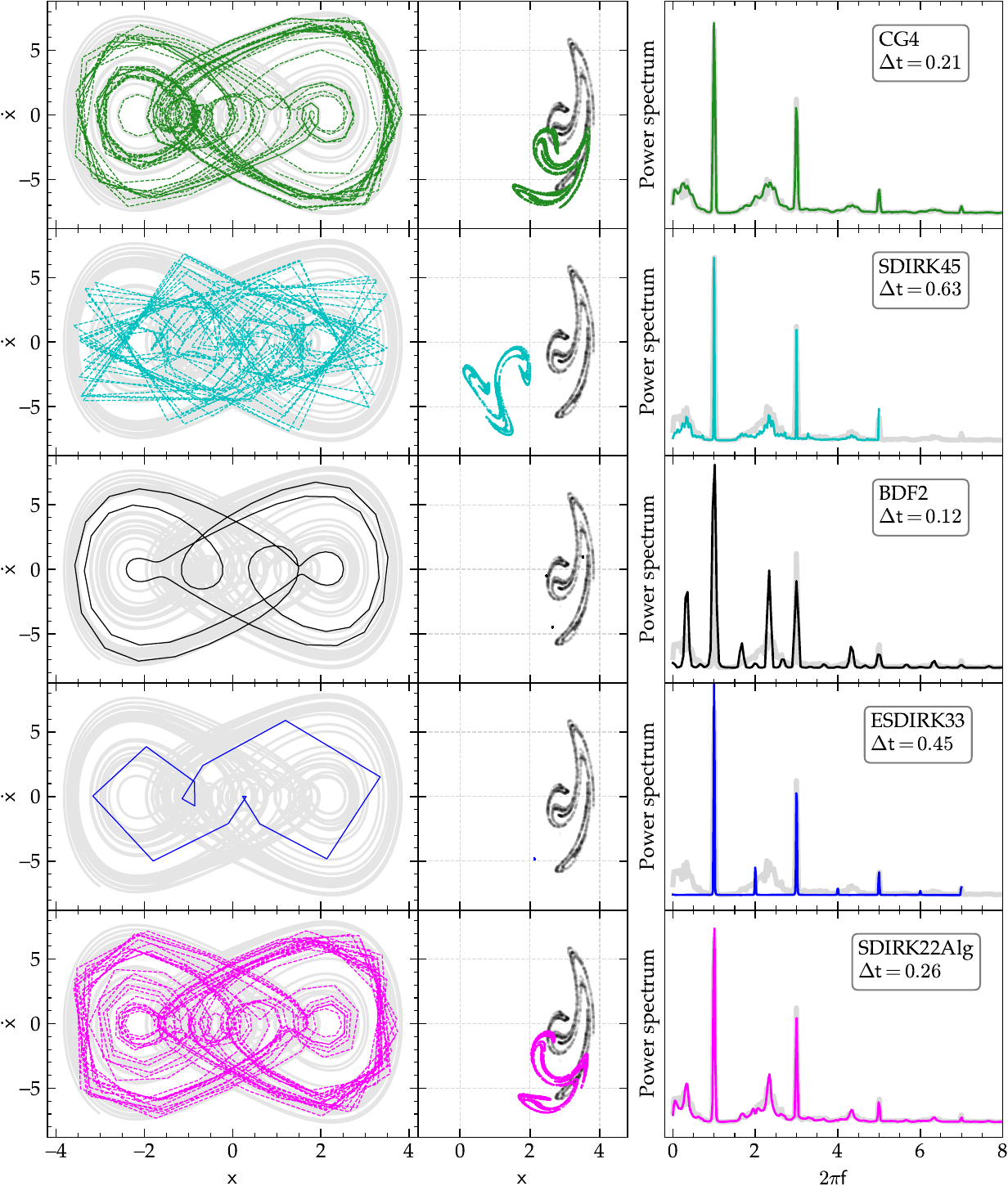}

     \caption{The trajectories of the chaotic attractor of the Duffing oscillator 
(left column), Poincare maps
(middle column), and the power spectra (right column) 
obtained with different implicit numerical schemes for the same computational 
efficiency ($\Delta t/<t_{\rm CPU}> \approx 450$). }
     \label{Fig: same efficiency 1}
\end{figure}

In all three rows of the right column of Fig.~\ref{Fig: lambdas vs time-step}, CG4 outperforms other schemes in terms of efficiency, while the third-order scheme shows the worst result. SDIRK22Alg shows better efficiency than SDIRK45 in the periodic regime (top row), performs on par with SDIRK45 in the main chaotic region (middle row), but fails to compete with SDIRK45 in predicting the standalone chaotic island (bottom row). BDF2, being among the cheapest implicit time integration schemes considered in this study, shows inferior results compared to SDIRK22Alg for this problem, because
it fails at reproducing $\lambda_1$ starting from smaller values of $\Delta t$
(the same behavior was seen in the Lorenz system).


\begin{figure}[h!]
    \centering
    \includegraphics[width=0.94\textwidth]{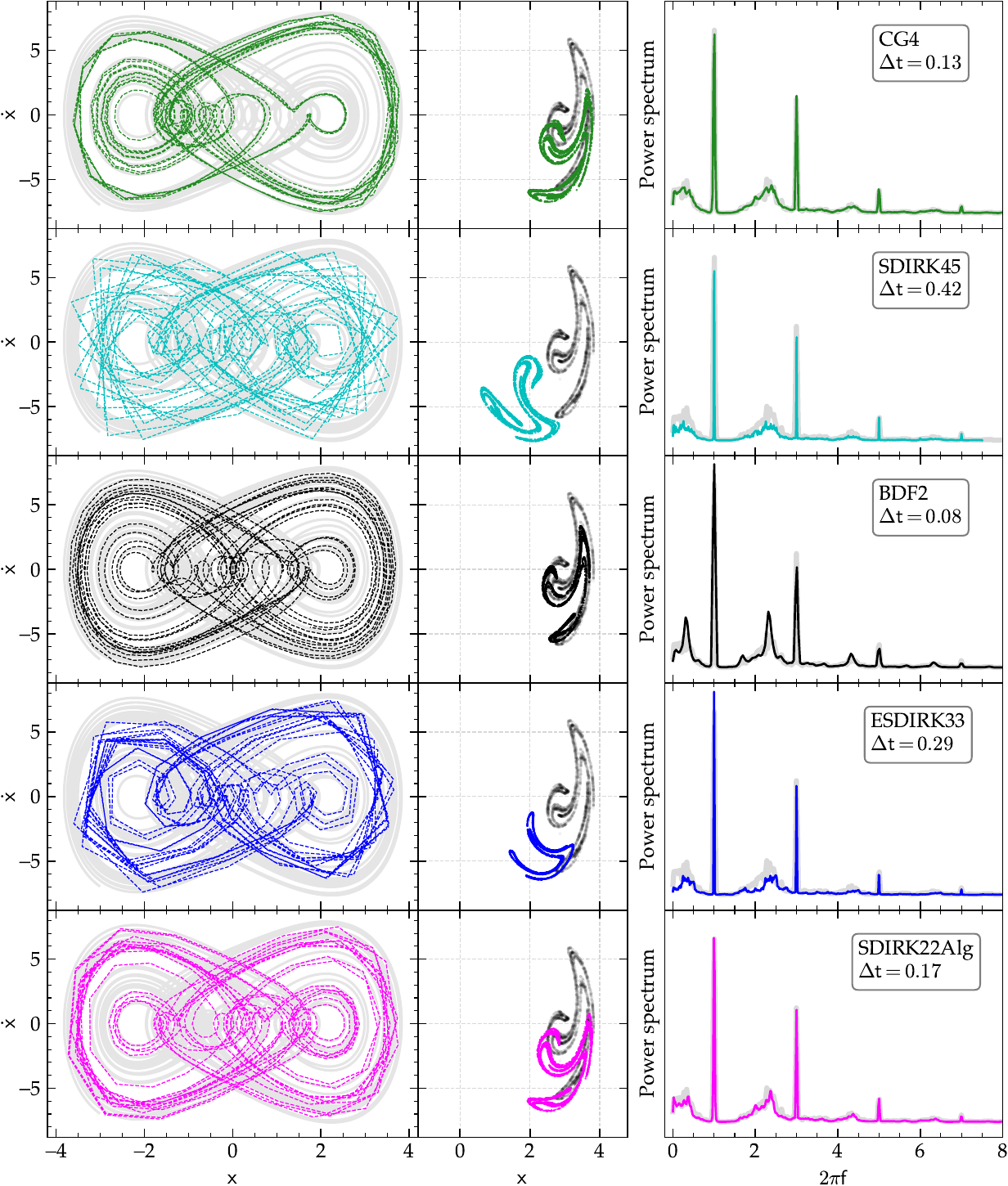}

     \caption{Same as Fig.~\ref{Fig: same efficiency 1}, but for $\Delta t/<t_{\rm CPU}> \approx 300$.}
     \label{Fig: same efficiency 2}
\end{figure}


Figs.~\ref{Fig: same efficiency 1} and~\ref{Fig: same efficiency 2} provide a more detailed comparison between the 
 different implicit schemes
for the same value of computational efficiency, and therefore the same total cost
of the simulations. In both figures, $\gamma=11$
(the same as in the two middle panels of Fig.~\ref{Fig: lambdas vs time-step}), 
so the correct behavior of the Duffing oscillator is chaotic. In all panels
of Figs.~\ref{Fig: same efficiency 1} and~\ref{Fig: same efficiency 2}, the results shown
in gray are the reference solutions obtained with the explicit RK4 scheme at 
$\Delta t = 10^{-3}$.
The quantities plotted in Figs.~\ref{Fig: same efficiency 1} and~\ref{Fig: same efficiency 2}, namely, the phase space
trajectories of the oscillator, the Poincare maps, and the power spectra,
were previously discussed in detail in \cite{Ueda1979}, and the 
RK4 simulation shown here agrees well with the 
published results for the same values of the parameters. The power spectra were obtained
using the function \texttt{signal.welch} from the SciPy package of Python, 
which is based on the Welch's method \cite{Welch1967}.

Fig.~\ref{Fig: same efficiency 1} corresponds to $\Delta t/<t_{\rm CPU}> \approx 450$, and the time-step sizes 
needed to achieve this efficiency are given in the rightmost column of the figure
for the different schemes\footnote{The time-step sizes were rounded to the closest 
integer fractions of the driving force period, $2\pi$, in order to be able to sample
the data at precisely one driving force period, which is important for getting
accurate Poincare maps.}. 
This specific value of $\Delta t/<t_{\rm CPU}>$
was chosen as the largest value at which more than one scheme can reproduce
the chaotic behavior of the attractor (above this value, only CG4 is able to
do it). For this reason, the simulations shown in Fig.~\ref{Fig: same efficiency 1} can be seen as the cheapest
possible for the Duffing oscillator.
Fig.~\ref{Fig: same efficiency 1} shows that CG4, SDIRK22Alg and SDIRK45, while obviously
running at coarse time-step sizes, reproduce the general shape of the
attractor well, and do especially well at reproducing the
power spectrum. BDF2 and ESDIRK33 are not able to reproduce the
behavior of Duffing oscillator for this value of computational 
efficiency.

Fig.~\ref{Fig: same efficiency 2} shows the same physical quantities as Fig.~\ref{Fig: same efficiency 1}, 
but for a different value of computational efficiency, namely,
$\Delta t/<t_{\rm CPU}> \approx 300$. This value was chosen as the largest one
at which the BDF2 scheme starts reproducing the chaotic behavior of the oscillator. 
In Fig.~\ref{Fig: same efficiency 2}, BDF2 demonstrates the best results among the implicit schemes,
especially in reproducing the Poincare map, 
in part because it has the smallest time-step size for the given
value of $\Delta t/<t_{\rm CPU}>$. 
However, all schemes shown in Fig.~\ref{Fig: same efficiency 2} are able 
to accurately reproduce the spectrum and the trajectory of the attractor.
Figs.~\ref{Fig: same efficiency 1} and~\ref{Fig: same efficiency 2} 
show that while BDF2 performs well, it is possible to 
run cheaper simulations of the Duffing oscillator 
that are still reasonably accurate using CG4, (E)SDIRK45,
or (E)SDIRK22(Alg).

\subsection{Kuramoto-Sivashinsky PDE}

The Kuramoto-Sivashinsky (KS) equation is probably best known for describing the
propagation of flame front instabilities \cite{Sivashinsky1977}, but it was shown to describe
the asymptotic dynamics of a wide range of
physical phenomena, including thin liquid film flows \cite{Papageorgiou1990}
and interfacial flows \cite{Hooper1985}, chemical reactions \cite{Kuramoto1976}, and
directional solidification \cite{Novick-Cohen1987}. The KS equation
studied here is formulated as
\begin{equation}
\label{KS_eq}
u_{t} = - uu_{x} - u_{xx} - u_{xxxx}\ ,
\end{equation}
but we should mention that there exists an alternative formulation
of this equation in terms
of a function $v$ such that $u=v_x$, which differs from Eq.~(\ref{KS_eq}) in
the non-linear term. The formulation given by Eq.~(\ref{KS_eq}) is commonly used in
fluid dynamics, because
the spatial average of the function $u$ remains constant in time, while in the
alternative formulation the spatial average of $v$ is subject to a drift, and the
equation has to be modified to prevent its growth \cite{Hyman1986}.

It is a well-known fact that the dynamics of the solution of the KS equation
depends on the spatial size of the domain, $L$
\cite{Manneville1985,Tajima2002,Edson2019}. For relatively small systems,
$L\lesssim13$, the solutions represent stable traveling waves, which start
developing
oscillations as the size of the system increases, and eventually transform into
a complex pattern of the so-called `spatio-temporal' or `weak' turbulence
\cite{Hyman1986a} (for a detailed explanation of the properties of weak
turbulence, see \cite{Manneville1990}). 
Even though there is no direct established link between the spatio-temporal
turbulence of the KS equation and hydrodynamical turbulence of the
Navier-Stokes equations, there is a number of similarities between the
two sets of equations, which indicate a possible connection and motivate 
further studies \cite{Holmes2012}. 
For example, the second order term in Eq.~(\ref{KS_eq}), $u_{xx}$, 
is destabilizing and acts as an energy source, while the fourth order
term, $u_{xxxx}$, provides a small-scale dissipation, acting as an energy sink, 
and both of these terms have analogs in the Navier-Stokes equations
\cite{Dankowicz1996}. The KS equation possesses translational and reflection
symmetries analogous to the ones of the Navier-Stokes equations for a
boundary layer \cite{Holmes2012}.
Thorough studies of the symmetries and steady state solutions of the KS
equation can be found, for example, in 
\cite{Michelson1986,Greene1988,Cvitanovic2010}.

Here, the KS equation were numerically solved for the initial conditions 
\begin{equation}
u(x,0) = \cos\left(\frac{x}{16}\right)\left(1+\sin\left(\frac{x}{16}\right)\right)\ ,
\end{equation}
previously studied in \cite{Xu2006,Kassam2005,Bhatt2021}. We employ
periodic boundary conditions in space, $u(L,t)=u(0,t)$, and use the spatial
domain $[0,L]=[0,32\pi]$. For this set of parameters, the solution develops
a chaotic pattern at time $t\approx 50$ and stays chaotic for the rest of
the evolution ($2\times 10^4$ in our simulations).

The spatial domain was uniformly discretized with $N=128$ points\footnote{We
performed some of the runs at a finer spatial 
resolution of $N=256$ points, and obtained the same results.}
and the sixth-order compact finite differencing was used to approximate the spatial 
derivatives \cite{Lele1992}. The time integration of all
terms on the right hand side of Eq.~(\ref{KS_eq}) was performed implicitly, using BDF2,
SDIRK45, SDIRK22Alg, and CG4 schemes. Purely implicit time integration
is not the most computationally
efficient way to solve the KS equation, and it is more common in the
literature to evolve the non-linear term explicitly, while using the implicit time
integration for the stiff linear term only (for a very large number of computational
methods used to integrate the KS equation, see the references 
in \cite{Bhatt2021}). However, since the goal of this study was to compare
the performance of different implicit numerical schemes in modeling
chaotic systems, we chose the simplest way to do the time integration,
as long as all schemes were put in equal conditions for the fair comparison.

\begin{figure}[h!]
    \centering
    \includegraphics[width=0.69\textwidth]{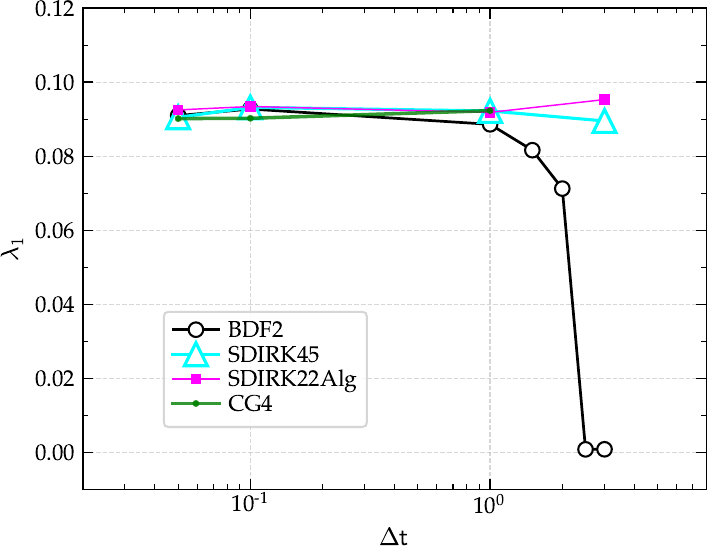}

     \caption{The leading Lyapunov exponents of the KS system obtained
with implicit schemes for different time-step sizes. The value of $\lambda_1$ computed with BDF2 deviates from
the correct value starting from $\Delta t > 1.0$, but the low computational cost
of this scheme keeps it competitive in terms of efficiency 
(see the text).}
     \label{Fig: lam1 KS}
\end{figure}

Fig.~\ref{Fig: lam1 KS} shows the leading Lyapunov exponents obtained at different time-step sizes. In the finest time resolution simulations,
we get 11 positive Lyapunov exponents, which is typical for the
relatively large size of the spatial domain used here, $L=32\pi \approx 100$.
It is common to use the so-called Kaplan-Yorke dimension \cite{Kaplan1979},
\begin{equation}
D_{KY} = j + \frac{\Sigma_{i=1}^{j}\lambda_i}{|\lambda_{j+1}|},
\end{equation}
where $j$ is the largest index such that $\Sigma_{i=1}^{j}\lambda_i\geq 0$,
to quantify the dimension of a chaotic attractor. At $\Delta t = 0.05$, we measure
$D_{KY} = 22.57$, which is consistent with the value $22.44$ reported in 
\cite{Edson2019} for $L=100$.
Fig.~\ref{Fig: lam1 KS} shows that CG4, SDIRK45 and SDIRK22Alg schemes consistently
predict the leading Lyapunov exponent to be $\lambda_1 \approx 0.09$ up to the largest time-step sizes
we could use in the simulations (SDIRK45 and SDIRK22Alg failed to converge for
$\Delta t=3.5$, while CG4 failed to converge already for $\Delta t=1.5$). This value of $\lambda_1$ agrees well 
with the value $\lambda_1=0.088$ reported in \cite{Edson2019} for 
$L=100$ and the periodic boundary conditions.
BDF2 did not fail to converge even at
the time-step sizes larger than $\Delta t=3.0$, but the value of $\lambda_1$
that it predicts starts to deviate from the correct value when $\Delta t \gtrsim 1.5$
and drops to nearly zero for $\Delta t \gtrsim 2.5$.

\begin{figure}[h!]
    \centering
    \includegraphics[width=0.99\textwidth]{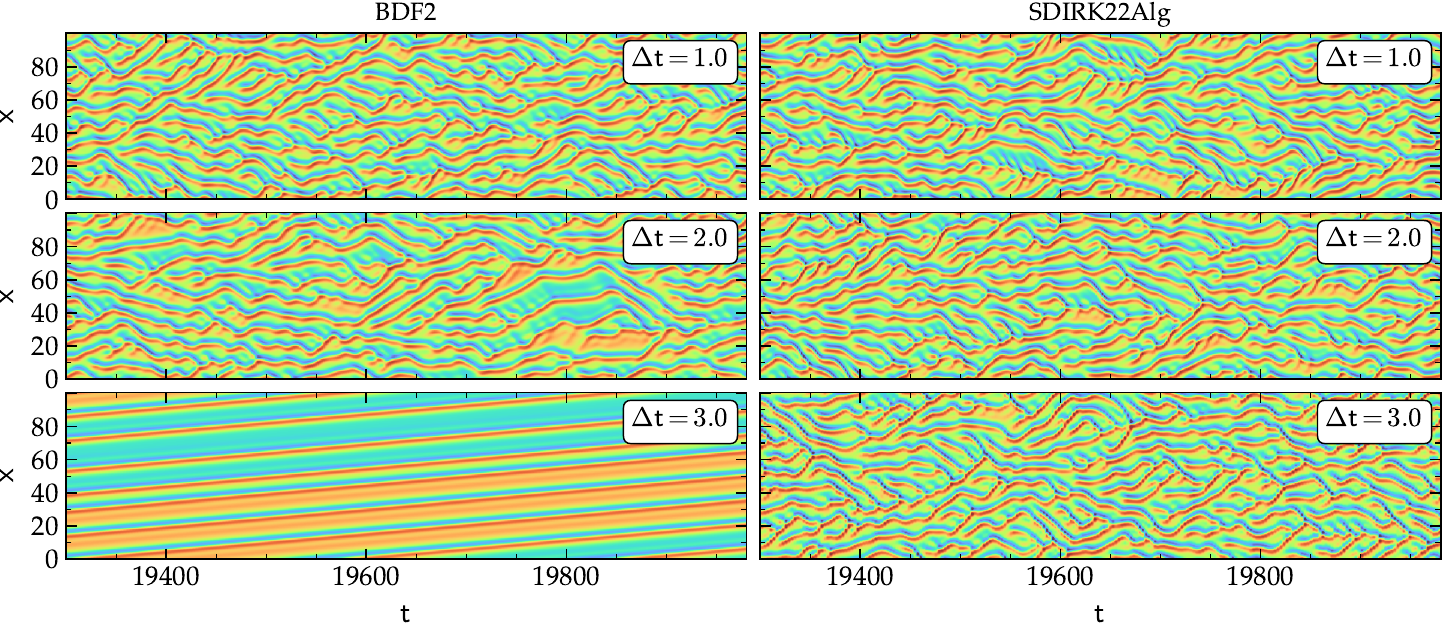}

     \caption{The color-coded values of the solution of the KS equation, $u(x,t)$, 
obtained with BDF2 and SDIRK22Alg schemes
for the different time-step sizes.
BDF2 (left column) shows relaminarization at the large values of $\Delta t$, 
while SDIRK22Alg (right column) predicts the same dynamics in all 
simulations. }
     \label{Fig: KS}
\end{figure}

Fig.~\ref{Fig: KS} shows the color-coded values of $u(x,t)$ and 
further demonstrates that the simulations done with BDF2 undergo 
relaminarization at large values of $\Delta t$, while SDIRK22Alg correctly predicts
the dynamics even at the largest time-step sizes. This starts to play an important
role when we compare the performance of the schemes in terms of their
computational efficiency. The total wall-clock times (indicative of the cost) of the
BDF2 simulations shown in Fig.~\ref{Fig: KS} are approximately $15$, $9$, and $5$ minutes
for $\Delta t = 1.0$, $2.0$, and $3.0$, while for SDIRK22Alg the corresponding
wall-clock times are $24.5$, $15$, and $11$ minutes\footnote{We emphasize that
the wall-clock times should be used only for the relative comparison between the
schemes, because their absolute values strongly depend on the hardware and 
software used in the simulations, as well as on the specific details
of the implementation, and the amount of work done by the script (for example,
the wall-clock times cited in this section include computation of the first 24 Lyapunov exponents,
as well as frequent output of the data).}. Therefore, the cheapest simulation that can
still relatively well capture the chaotic dynamics of the KS system is the BDF2 
simulation at the time-step size $\Delta t = 2.0$. 
However, $\lambda_1$ measured in that simulation is only $\sim 0.07$ versus
the correct value of $\sim 0.09$, and if the accuracy is relevant for the intended application 
of the simulation, the SDIRK22Alg run at $\Delta t = 3.0$ may be preferable.

CG4 and SDIRK45 predict the same dynamics of the KS system 
as shown in Fig.~\ref{Fig: KS} for SDIRK22Alg, but for this problem they are not
competitive in terms of the efficiency. The CG4 simulation at $\Delta t = 1.0$
is about four times more expensive than the corresponding BDF2 simulation, and
the lack of stability of the CG4 scheme limits its ability to work at larger
time-step sizes.
All SDIRK45 runs were $2$ to $2.5$ times slower than the SDIRK22Alg runs
for the same $\Delta t$, which is expected based on the number of stages
in these schemes. To investigate the performance of the other second-order 
schemes, we run additional simulations at $\Delta t = 1.0$, $2.0$,
and $3.0$, using SDIRK22 and ESDIRK22. For the same value of $\Delta t$,
the wall-clock times of all second-order simulations differed by less than $10\%$,
and both SDIRK22 and ESDIRK22 correctly predicted $\lambda_1$ at large
time-step sizes, but the maximum permissible $\Delta t$ values were smaller than that of SDIRK22Alg. The coarsest run that we could perform with 
SDIRK22 was at $\Delta t\approx 2.0$, while for ESDIRK22 it was at
$\Delta t\approx 1.0$. Therefore, better stability properties of SDIRK22Alg
played a significant role in its computational efficiency in this test.


\begin{figure}[h!]
    \centering
    \includegraphics[width=0.99\textwidth]{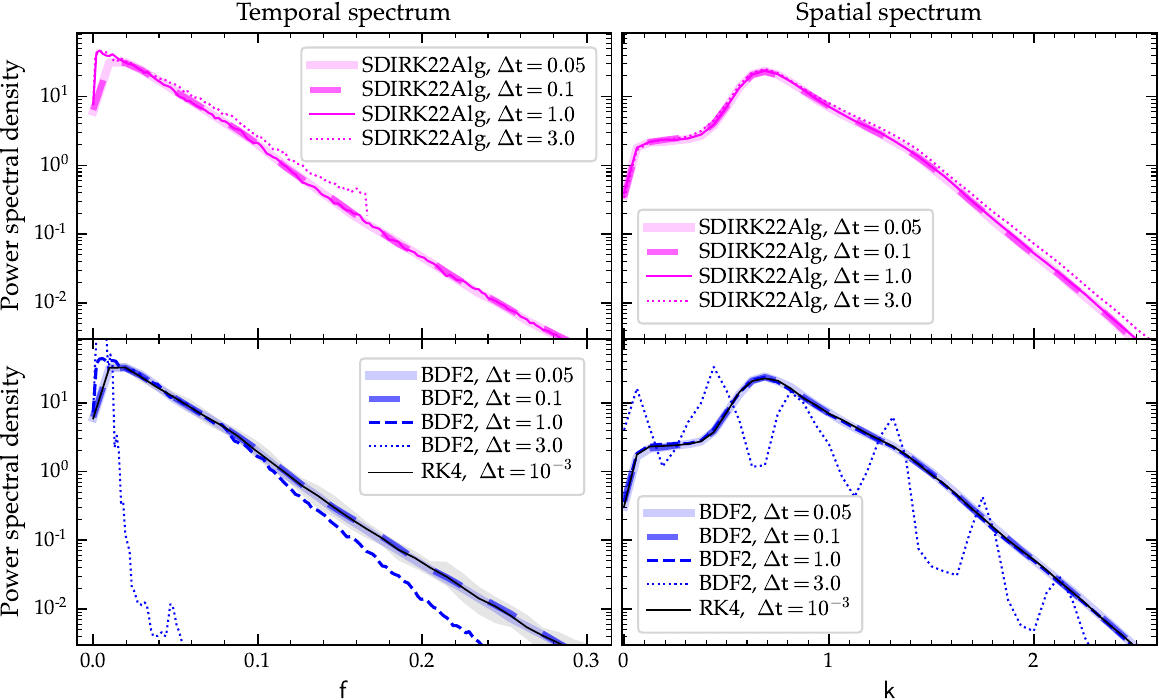}

     \caption{Temporal (left column) and spatial (right column) power spectral
densities of the solution of the KS equation obtained with 
SDIRK22Alg (top panels) and BDF2 (bottom
panels). The bottom panels also show the results obtained with the
explicit RK4 integration at $\Delta t = 10^{-3}$. }
     \label{Fig: power spectra}
\end{figure}

Finally, Fig.~\ref{Fig: power spectra} shows the power spectral densities obtained from the
BDF2 and SDIRK22Alg simulations at different time-step sizes. For the temporal
spectra (left column), the values of $u$ are collected as a function of time at ten coordinate
locations distributed uniformly across the spatial domain. The
power spectral densities are computed separately at each location using Welch's method,
and the averaged spectral densities are shown in Fig.~\ref{Fig: power spectra}. 
As in Figs.~\ref{Fig: lam1 KS} and~\ref{Fig: KS}, SDIRK22Alg produces the same result for 
all time-step sizes used in the simulations (the same is true for CG4 and
SDIRK45, which are not shown in Fig.~\ref{Fig: power spectra}). BDF2, on the other
hand, starts to deviate from the converged spectrum at $\Delta t \gtrsim 1.0$.
To verify the results, we performed an explicit RK4 simulation of the KS
system at the time-step size $\Delta t=10^{-3}$, 
and its temporal spectrum is shown in the bottom left panel
of Fig.~\ref{Fig: power spectra} with a black curve. The gray region surrounding the
RK4 curve fills the space covered by all spectra computed
at different spatial locations and gives a sense of the errorbar admitted when
extracting the averaged spectrum in this way. All spectra in the left
column of Fig.~\ref{Fig: power spectra} have a scatter similar to the one of RK4, but
we do not show it for each simulation to avoid cluttering the plot.

The right column of Fig.~\ref{Fig: power spectra} shows the spatial spectra computed from the
snapshots of $u$ taken at constant values of time and averaged 
across all times.
The spatial spectra for the KS system were previously published in 
\cite{Toh1987,Dankowicz1996,Holmes2012}, and our simulations agree well with
the published results. In particular, the characteristic maximum in the
spatial spectrum at $k\approx1/\sqrt{2}\approx 0.7$ indicates the
size of the most energetic waves in the solution.
All spectra shown in the right column of Fig.~\ref{Fig: power spectra} demonstrate an excellent convergence
apart from the BDF2 spectrum at $\Delta t = 3.0$, due to the relaminarization.

Among the chaotic systems considered in this study, the Kuramoto-Sivashinsky system is the closest in terms of complexity to the hydrodynamical turbulence, and the results obtained for this system help us draw final conclusions about the effectiveness of different implicit schemes for modeling dynamical systems. The spectra presented in Fig.~\ref{Fig: power spectra} for all but the very largest time-step sizes are accurate enough to be suitable for many engineering applications, which tells that performing the simulations close to the stability limits of the implicit schemes may produce satisfactory results. At the same time, Fig.~\ref{Fig: lam1 KS} shows that more costly, fourth-order implicit methods do not allow an increase in the maximum time-step size compared to the cheaper second-order schemes. This speaks in favor of using second-order numerical schemes when solving Navier-Stokes equations numerically in the turbulent regime.

\section{Conclusions}

Chaotic systems are characterized, among other features, by the presence of
dense periodic phase space orbits \cite{Devaney2021}, which means that the time-step
sizes required to accurately represent their dynamics 
in numerical simulations are restricted from above
by the duration of their (quasi-) periods. For the systems considered here, we find
that this requirement is more stringent than the stability limits of the schemes,
meaning that all implicit schemes fail at approximately same time-step sizes
once the time-step size reaches a significant fraction of the period of the system.
Unlike in laminar systems, in chaotic systems the implicit schemes are able
to provide only modest gain in terms of the time-step size when compared to explicit
fourth-order Runge-Kutta method. For this reason, if specifics of the problem
force one to use implicit time integration methods, the computational cost per time-step
becomes the leading factor in choosing the scheme.

This study shows that at large time-step sizes, the third- and fourth-order schemes, due to their larger cost, do not 
provide significant benefits in performance compared to the second-order
schemes, and the third-order schemes perform worse than the fourth-order
ones. An interesting exception was the CG4 scheme, which 
outperformed all other implicit schemes in terms of the computational efficiency for the
Lorenz system and Duffing oscillator. However, the lack of \textit{L}-stability will
likely restrict the use of this scheme in hydrodynamical turbulence simulations, and
indeed, it could not compete with a more stable second-order scheme
in terms of efficiency at solving the Kuramoto-Sivashinsky equation.
In addition, the benefits of CG4 can be eliminated by using a machine with different hardware/software, and we have seen it performing worse on a laptop computer. 

BDF2, being among the cheapest and easiest to implement schemes considered in our study, 
represents an acceptable choice for modeling chaotic systems. However, in all
our experiments, BDF2 showed more restricted time-step size requirements for physical accuracy compared to other schemes,
while preserving numerical convergence. When solving the
Kuramoto-Sivashinsky equation, it was the only scheme that produced
an incorrect value of the leading Lyapunov exponent for the time-step sizes considered and eventually
relaminarized the system, while other schemes predicted the Lyapunov
exponent correctly until the time-step size was so large the Newton-Raphson solver could no longer converge. For this reason, 
when working with BDF2, it is imperative to test the physical results for
convergence using more than one time-step size.

Summarizing, we have found that the second-order (E)SDIRK-type
schemes work best for the chaotic systems among the implicit methods.
Although SDIRK22, ESDIRK22 and SDIRK22Alg showed almost identical
performance in modeling the Lorenz system, in the Kuramoto-Sivashinsky test
the additional stability properties of SDIRK22Alg allowed it to reach larger
time-step sizes and therefore to be the most computationally efficient.
However, the superiority of SDIRK22Alg over the other \nth{2}-order methods may be predicated on deep convergence of the non-linear residuals, and practical application may favor ESDIRK22 for its higher stage order.


This study was mainly concerned with testing the limits of the implicit schemes at large time-step sizes, which are expected to be most practical, and the conclusions collected here apply to that regime. As the accuracy requirements tighten and force one to use smaller timesteps, the fourth-order methods become more effective, although, depending on the details of the problem, in that regime the preference may be given back to the explicit schemes. Since each problem would require individual approach and it is hard to draw general conclusions, we leave it up to reader to evaluate pros and cons of using higher-order methods in their application at small time-step sizes.

\section*{Funding Sources}

This material is based in part upon work supported by the Air Force Office of Scientific Research under award number FA9550-22-1-0108. Any opinions, findings, and conclusions or recommendations expressed in this material are those of the author(s) and do not necessarily reflect the views of the United States Air Force.

\bibliography{references}

\appendix

\section{Butcher Tableaux for Specific (E)SDIRK Methods}

The construction of (E)SDIRK schemes with varying numbers of stages, orders of accuracy, and stability properties often leads to free coefficients \cite{Kennedy2016}. The following describes the specific Butcher tableaux used in this work.

\begin{table}[!h]
	\renewcommand\arraystretch{1.4}
	\centering
	$
	\begin{array}{c|ccccc}
	1 - \frac{\sqrt{2}}{2} & 1 - \frac{\sqrt{2}}{2} \\
	1                      & \frac{\sqrt{2}}{2} & 1 - \frac{\sqrt{2}}{2} \\ \hline
	                       & \frac{\sqrt{2}}{2} & 1 - \frac{\sqrt{2}}{2}
	\end{array}
	$
	\label{sdirk22_tableau}
	\caption{SDIRK22. \cite{Kennedy2016}}
\end{table}

\begin{table}[!h]
	\renewcommand\arraystretch{1.2}
	\centering
	$
	\begin{array}{c|ccccc}
	0.4358665215 & 0.4358665215 \\
	0.7179332608 & 0.2820667392 & 0.4358665215 \\
	1            & 1.208496649  & -0.644363171 & 0.4358665215 \\ \hline
	             & 1.208496649  & -0.644363171 & 0.4358665215
	\end{array}
	$
	\label{sdirk33_tableau}
	\caption{SDIRK33. (Note: the coefficients for this scheme are reported in Kennedy and Carpenter \cite{Kennedy2016} as functions of the diagonal value rather than rational coefficients.)}
\end{table}

\begin{table}[!h]
	\renewcommand\arraystretch{1.4}
	\centering
	$
	\begin{array}{c|ccccc}
	\frac{1}{4}   & \frac{1}{4} \\
	\frac{3}{4}   & \frac{1}{2}      & \frac{1}{4} \\
	\frac{11}{20} & \frac{17}{50}    & -\frac{1}{25}     & \frac{1}{4} \\
	\frac{1}{2}   & \frac{371}{1360} & -\frac{137}{2720} & \frac{15}{544} & \frac{1}{4} \\
	1             & \frac{25}{24}    & -\frac{49}{48}    & \frac{125}{16} & -\frac{85}{12} & \frac{1}{4} \\ \hline
	              & \frac{25}{24}    & -\frac{49}{48}    & \frac{125}{16} & -\frac{85}{12} & \frac{1}{4}
	\end{array}
	$
	\label{sdirk45_tableau}
	\caption{SDIRK45. \cite{Kennedy2016}}
\end{table}


\begin{table}[!h]
	\renewcommand\arraystretch{1.2}
	\centering
	$
	\begin{array}{c|ccccc}
        0 & 0 \\
	2 - \sqrt{2} & 1 - \frac{\sqrt{2}}{2} & 1 - \frac{\sqrt{2}}{2} \\
	1            & \frac{\sqrt{2}}{4}  & \frac{\sqrt{2}}{4} & 1 - \frac{\sqrt{2}}{2} \\ \hline
	             & \frac{\sqrt{2}}{4}  & \frac{\sqrt{2}}{4} & 1 - \frac{\sqrt{2}}{2}
	\end{array}
	$
	\label{esdirk22_tableau}
	\caption{ESDIRK22. \cite{Kennedy2016}}
\end{table}

\begin{table}[!h]
	\renewcommand\arraystretch{1.4}
	\centering
	$
	\begin{array}{c|cccc}
	0   & 0 \\
	\frac{1767732205903}{2027836641118}   & \frac{1767732205903}{4055673282236}      & \frac{1767732205903}{4055673282236} \\
	\frac{3}{5} & \frac{2746238789719}{10658868560708}    & -\frac{640167445237}{6845629431997}     & \frac{1767732205903}{4055673282236} \\
	1   & \frac{1471266399579}{7840856788654} & -\frac{4482444167858}{7529755066697} & \frac{11266239266428}{11593286722821} & \frac{1767732205903}{4055673282236} \\
	\hline
    & \frac{1471266399579}{7840856788654} & -\frac{4482444167858}{7529755066697} & \frac{11266239266428}{11593286722821} & \frac{1767732205903}{4055673282236}
	\end{array}
	$
	\label{esdirk33_tableau}
	\caption{ESDIRK33. \cite{Bijl2002}}
\end{table}

\begin{table}[!h]
	\renewcommand\arraystretch{1.4}
	\centering
	$
	\begin{array}{c|cccccc}
	0   & 0 \\
	\frac{1}{2}   & \frac{1}{4}      & \frac{1}{4} \\
	\frac{83}{250} & \frac{8611}{62500}    & -\frac{1743}{31250}     & \frac{1}{4} \\
	\frac{31}{50}   & \frac{5012029}{34652500} & -\frac{654441}{2922500} & \frac{174375}{388108} & \frac{1}{4} \\
	\frac{17}{20}             & \frac{15267082809}{155376265600}    & -\frac{71443401}{120774400}    & \frac{730878875}{902184768} & \frac{2285395}{8070912} & \frac{1}{4} \\
	1             & \frac{82889}{524892}    & 0   & \frac{15625}{83664} & \frac{69875}{102672} & -\frac{2260}{8211} & \frac{1}{4} \\
	\hline
	              & \frac{82889}{524892}    & 0   & \frac{15625}{83664} & \frac{69875}{102672} & -\frac{2260}{8211} & \frac{1}{4}
	\end{array}
	$
	\label{esdirk45_tableau}
	\caption{ESDIRK45. \cite{Bijl2002}}
\end{table}


\end{document}